\documentclass[aps,prx,twocolumn,superscriptaddress,letterpaper]{revtex4-1}

\usepackage{amssymb}
\usepackage{amsmath}
\usepackage{bm}
\usepackage{graphicx}
\usepackage{color}
\usepackage{epsfig}
\usepackage{ulem}
\usepackage{siunitx}
\usepackage{float}
\usepackage{enumerate}
\usepackage{hyperref}
\hypersetup{colorlinks=true, citecolor=blue, urlcolor=blue, linkcolor=blue}

\definecolor{xred}{rgb}{0.6, 0.1, 0.15}

\begin{document}

\title{Non-Hermitian Photonic Lattices: tutorial}

\author{Qiang Wang}
\affiliation{Division of Physics and Applied Physics, School of Physical and Mathematical Sciences, Nanyang Technological University,
Singapore 637371, Singapore}

\author{Y. D. Chong}
\email{yidong@ntu.edu.sg}
\affiliation{Division of Physics and Applied Physics, School of Physical and Mathematical Sciences, Nanyang Technological University,
Singapore 637371, Singapore}
\affiliation{Centre for Disruptive Photonic Technologies, Nanyang Technological University, Singapore, 637371, Singapore}

\begin{abstract}
  Non-Hermitian photonic lattices combine the peculiar consequences of energy non-conservation with the physics of bandstructures, giving rise to a variety of exotic properties not found in conventional materials or photonic metamaterials.  In this tutorial, we introduce the key concepts in the design and implementation of non-Hermitian photonic lattices, including the general features of non-Hermitian lattice Hamiltonians and their bandstructures, the role of non-Hermitian lattice symmetries, and the topological chracterization of non-Hermitian bandstructures.  We survey several important non-Hermitian lattice designs, as well as the photonic platforms on which they can be realized.  Finally, we discuss the possibilities for future developments in the field.
\end{abstract}

\maketitle

\newpage
\section{Introduction}

The modern science of photonics has at its disposal extraordinary capabilities for fabricating structures to manipulate the flow of light.  In seeking out new ways to exploit these capabilities, photonics researchers have often found it useful to draw lessons from quantum theory.  Many insights and concepts originally formulated for quantum wavefunctions have been fruitfully applied to electromagnetic fields; examples include, but are not limited to, the invention of photonic crystals as electromagnetic analogues of electronic band insulators \cite{John87, Yablonovitch87}; the development of parity/time-reversal (PT) symmetric photonics based on a hypothetical non-energy-conserving formulation of quantum mechanics \cite{bender1998real, ElGanainy07, bender2007making, feng2017non, christodoulides2018parity}; and the development of photonic devices that mimick topological insulators and other topological phases of matter \cite{lu2014topological, ozawa2019topological}.  Over the past decades, photonic platforms have proven indispensable for realizing and investigating many such phenomena, including some that are hard or impossible to access in the analogous quantum mechanical context.  This has enabled a virtuous cycle of interactions between fundamental theory and practical experimentation.

The topic of this tutorial lies at one such theoretical and experimental frontier: the development of non-Hermitian photonic lattices.  These are photonic systems that (i) possess a lattice structure, e.g.~a photonic crystal or waveguide array, and (ii) deliberately violate Hermiticity, e.g.~through the incorporation of gain or loss.  By combining these two features, certain basic aspects of wave propagation can be altered in remarkable and surprising ways, with consequences that are still being actively explored.

It is a well-known fact, originating in condensed matter physics but since extending to other disciplines, that wave propagation in a periodic lattice is governed by a bandstructure \cite{kittel1996introduction, joannopoulos2011photonic}.  Bandstructures with different properties can be realized with various lattice configurations (e.g., different choices of lattice symmetry).  The bandstructure concept was originally formulated for Hermitian systems (i.e., those that obey the conservation of energy in detail).  This assumption is especially deeply embedded in the theory of band topology, which classifies materials into distinct ``topological phases'' based on whether their bands can be smoothly deformed into each other \cite{hasan2010colloquium, BansilReview2016}.  In the real world, of course, violations of Hermiticity are ubiquitous, such as the loss of energy caused by particles leaking out into the environment \cite{lindblad1976generators, bender2007making, rotter2009non}.  Nonetheless, lattices and their bandstructures have traditionally been studied in the Hermitian limit, with losses and other sources of non-Hermiticity treated as secondary issues.

In the 2000s, researchers began investigating the properties of non-Hermitian bandstructures in earnest, starting with the study of PT symmetric optical waveguide arrays \cite{ElGanainy07, makris2008beam, klaiman2008visualization, longhi2009bloch, schomerus2013topologically}, and continuing into lattices obeying other non-Hermitian symmetries \cite{bender2007making, feng2017non, el2018non, ozdemir2019parity, krasnok2021parity}.  It was found that non-Hermitian lattices can support propagating eigenmodes with real or complex eigenvalues, or a mix of the two.  Their energy bands can take on unusual forms, e.g.~coalescing at ``exceptional points'' (EPs) or along exceptional curves/surfaces, a phenomenon completely different from the band degeneracies found in Hermitian bandstructures \cite{heiss2001chirality, szameit2011p, heiss2012physics}.  These theoretical advances were closely accompanied by the experimental realization of non-Hermitian lattices on multiple photonic platforms, including optical waveguides \cite{guo2009observation, ruter2010observation} and microring resonators arrays \cite{peng2014parity, chang2014parity}.  Several novel device functionalities were demonstrated, such as unidirectional invisibility \cite{lin2011unidirectional, ge2012conservation, mostafazadeh2013invisibility, sounas2015unidirectional}, negative refraction \cite{fleury2014negative}, and stabilizing laser modes \cite{zhao2018topological}.  Researchers also found many functionalities that can be accessed by non-Hermitian systems not configured as lattices \cite{bender2007making}, such as single- or few-mode resonator systems that act as coherent perfect absorbers \cite{chong2010coherent, longhi2010pt, baranov2017coherent}, or aim to use EPs to enhance optical sensing \cite{wiersig2014enhancing, wiersig2016sensors, hodaei2017, chen2017exceptional, zhong2019sensing, hokmabadi2019non, park2020symmetry, qin2021exceptional}; these lie outside the scope of our discussion.

More recently, the topology of non-Hermitian bands has been the focus of much attention.  Standard theories of band topology \cite{hasan2010colloquium, BansilReview2016, davis2021photonic}, which classify bandstructures into discrete topological classes that predict the existence of ``topologically protected'' states on their boundaries, rest on the assumption that the system is Hermitian.  Most advances in the field of topological photonics \cite{lu2014topological, ozawa2019topological} have also been based on this Hermitian framework.  Theoretical efforts to reformulate the band topology concept for non-Hermitian systems \cite{rudner2009topological, liang2013topological, shen2018topological, ghatak2019new, kawabata2019symmetry, torres2019perspective, bergholtz2021exceptional, zhang2022review} have yielded several surprises, such as the ability of non-Hermitian degrees of freedom (e.g., gain and loss) to drive systems across topological phase transitions \cite{zeuner2015observation, takata2018photonic, longhi2021phase, liu2020generalized}, and the existence of special boundary states not tied to standard Hermitian topological invariants \cite{lee2016anomalous, leykam2017edge, luo2019higher}.  Photonic lattices have been platforms of choice for the experimental exploration of these concepts  \cite{regensburger2012parity, zhen2015spawning, xu2016experimental, weimann2017topologically, xiao2017observation, zhao2018topological}.  There is also increasing interest in the ``non-Hermitian skin effect'', whereby certain non-Hermitian lattices' eigenstates condense unexpectedly onto their boundaries; this phenomenon has been linked to band topology in some, though not all, models \cite{bergholtz2021exceptional, zhang2022review, ding2022non}, and has potential applications for concentrating light \cite{weidemann2020topological} and the maintenance of single-mode laser operation \cite{ZhuAnomalous2022}.

In this tutorial, we will explain the key concepts behind non-Hermitian photonic lattices, and discuss how they are theoretically designed and realized in experiments.  We begin in Sec.~\ref{nonherm-hams} by analyzing a simple two-mode non-Hermitian Hamiltonian and its complex energy levels.  In Sec.~\ref{sec:Symmetry}, we discuss how the symmetries of non-Hermitian systems can give rise to special behaviors, and survey the most prominent of these symmetries.  In Sec.~\ref{sec:lattices}, we provide an overview of some important non-Hermitian lattice models in 1D and higher dimensions.  In Sec.~\ref{sec:topology}, we discuss how the topology of non-Hermitian bands can be defined, and the physical implications.  In Sec.~\ref{sec:platform}, we review the experimental platforms that can be used to implement non-Hermitian photonic lattices.  In Sec.~\ref{sec:outlook}, we provide an outlook on the field, and suggest some areas in which future advances might be made.

\section{Non-Hermitian Hamiltonians}
\label{nonherm-hams}

\subsection{A Simple Non-Hermitian Hamiltonian}
\label{sec:model}

\begin{figure*}
\centering
\includegraphics[width=0.98\textwidth]{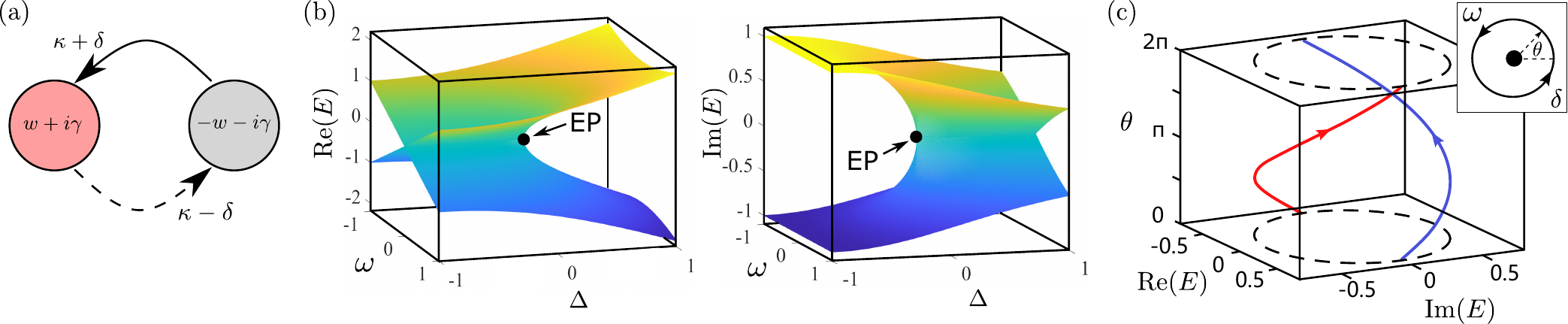}
\caption{(a) Schematic of a pair of coupled resonators with frequency detuning $\pm \omega$, gain/loss rates $\pm \gamma$, and inter-resonator couplings $\kappa \pm \delta$.  (b) Real and imaginary parts of the eigenenergies versus $\omega$ and $\Delta$, for $\gamma =1$ and $\kappa = \gamma+\Delta$.  The order-2 exceptional point (EP$_2$) is indicated by a black dot.  (c) Plot of the complex eigenenergy trajectory when encircling the EP$_2$, along the parametric curve $\omega=r\sin\theta$ and $\kappa=\gamma+r\cos\theta$, where $r = 0.2$.  
}
\label{fig:EP2}
\end{figure*}

We begin by discussing a simple two-mode Hamiltonian, which can help to
illustrate several key features of non-Hermitian Hamiltonians.  Consider the $2\times 2$ matrix
\begin{equation}
\mathcal{H}=\begin{pmatrix} \omega+i\gamma & \kappa+\delta \\ 
\kappa-\delta & -\omega-i\gamma \end{pmatrix},
\label{H2by2}
\end{equation}
where $\omega, \gamma, \kappa, \delta \in \mathbb{R}$.

Such Hamiltonians can arise in photonics in different ways.  For instance, in the framework of coupled-mode theory \cite{haus1991coupled, fan2003cmt}, $\mathcal{H}$ may describe a system of two coupled resonators, as depicted in Fig.~\ref{fig:EP2}(a).  Suppose the resonators host one resonant mode each, with complex electric field profiles $\mathbf{E}_1(\mathbf{r})$ and $\mathbf{E}_2(\mathbf{r})$.  When the resonators are decoupled, let the modes have oscillation frequencies $\omega_0 \pm \omega$ (where $\omega_0 \in \mathbb{R}$ is some central frequency) and decay rates $\pm\gamma$ (where a negative decay rate means the mode experiences amplification, or gain, over time).  If we write the total field using the ansatz
\begin{equation}
  \mathbf{E}(\mathbf{r},t) = \left[\psi_1(t) \mathbf{E}_1(\mathbf{r}) + \psi_2(t) \mathbf{E}_2(\mathbf{r})\right] \, e^{-i\omega_0 t},
\end{equation}
then the time-dependent modal coefficients obey
\begin{equation}
  i \frac{d}{dt} \begin{pmatrix} \psi_1 \\ \psi_2 \end{pmatrix}
  = \begin{pmatrix} \omega + i\gamma & 0 \\ 0 & -\omega - i\gamma
  \end{pmatrix} \begin{pmatrix} \psi_1 \\ \psi_2 \end{pmatrix},
  \label{modalcoeff}
\end{equation}
which has solutions $\psi_1(t) = \psi_1(0) \exp(-i\omega t) \exp(-\gamma t)$ and
$\psi_2(t) = \psi_1(0) \exp(i\omega t) \exp(\gamma t)$.  We can then couple the two resonators, in order to introduce off-diagonal elements into the matrix in \eqref{modalcoeff} and thereby bring it into the form of Eq.~\eqref{H2by2}.  For now, we will gloss over the details of how these matrix elements might be determined.  As coupled-mode theory is built out of approximations to the underlying Maxwell equations, its predictions are accurate only within a certain regime of validity (e.g., we typically require $\omega \ll \omega_0$, so that the modal coefficients represent slowly-varying envelope functions) \cite{haus1991coupled, fan2003cmt}.  Moreover, the parameters of the theory ($\omega$, $\gamma$, etc.)~are generally not physically independent, as variations in physical parameters (e.g., the refractive index of the resonators) can alter them simultaneously.  Other ways of deriving Hamiltonians for photonic systems are also similarly subject to various limitations (see Sec.~\ref{sec:platform}).

The Hamiltonian \eqref{H2by2} is non-Hermitian ($\mathcal{H}\neq \mathcal{H}^\dagger$) when $\gamma \neq 0$ and/or $\delta \neq 0$.  If $\gamma \ne 0$, one mode experiences gain and the other has an equal rate of loss.  If $\delta \ne 0$, the couplings in opposite directions have different magnitudes \cite{hatano1996localization, hatano1997vortex, yao2018edge} and are said to be nonreciprocal or asymmetric (see Sec.~\ref{sec:topology}).  We shall see that these two sources of non-Hermiticity---gain/loss and nonreciprocal couplings---have distinct physical consequences.

The eigenvalues and (unnormalized) right eigenvectors of $\mathcal{H}$ are
\begin{align}
E_{\pm}&=\pm \sqrt{(\omega+i\gamma)^2+\kappa^2-\delta^2} \label{H2by2eval}
\\
|\psi_{\pm}^R\rangle&=\begin{pmatrix} \kappa+\delta \\ E_\pm - \omega-i\gamma \end{pmatrix}.
\label{H2by2evec}
\end{align}
As we vary the parameters of $\mathcal{H}$ (e.g., $\omega$ and $\kappa$), these form a continuous bandstructure.  We will henceforth refer to the eigenvectors as ``eigenstates'', to emphasize their physical significance.    For each right eigenstate, there is also a left eigenstate with the same eigenvalue: $\langle\psi_{\pm}^L| \mathcal{H} = E_\pm \langle \psi_{\pm}^L|$.  In bra (row vector) form,
\begin{equation}
  \langle\psi_{\pm}^L| = \begin{pmatrix} \kappa - \delta,\;\; & E_\pm - \omega - i \gamma
  \end{pmatrix}.
\end{equation}
And in ket (column vector) form,
\begin{equation}
  |\psi_{\pm}^L\rangle = \begin{pmatrix} \kappa - \delta \\ E_\pm^* - \omega + i \gamma
  \end{pmatrix}.
\end{equation}
In general, $|\psi_\pm^L\rangle \ne |\psi_\pm^R\rangle$, which is typical of non-Hermitian matrices.  We sometimes omit the ``R'' superscript and denote right eigenstate as $|\psi_{\pm}\rangle$; likewise, ``eigenstate'' refers to a right eigenstate unless otherwise specified.

The behavior of the eigenvalue spectrum and the eigenstates falls into several distinct cases:

\textbf{Case I}: $\omega=\delta=0$, with $\gamma \ne 0$.  Here, the non-Hermiticity originates from the gain and loss of the two modes, which results in the breaking of \textit{time-reversal symmetry}: $[\hat{T},\mathcal{H}] \neq 0$, where $\hat{T}$ is the complex conjugation operation (an antilinear operator). However, $\mathcal{H}$ still satisfies the symmetry $[\hat{P}\hat{T}, \mathcal{H}] = 0$, where $\hat{P}=\sigma_1$ is the first Pauli matrix; this is called parity/time-reversal (PT) symmetry (see Section~\ref{sec:pt}).

If $|\gamma|<\kappa$, the eigenvalues $E_{\pm}=\pm \sqrt{\kappa^2-\gamma^2}$ are real, like those of a Hermitian system, and the system is said to be ``PT-unbroken'' \cite{bender2007making}.  If $|\gamma|>\kappa$, $E_{\pm}$ are both purely imaginary and the system is ``PT-broken''.

At the critical point $|\gamma|=|\kappa|$, we have $E_+ = E_- = 0$.  This eigenvalue degeneracy is an exceptional point (EP) \cite{heiss2001chirality, heiss2012physics, mailybaev2005geometric, feng2017non, el2018non, miri2019exceptional, ozdemir2019parity, ashida2020non, wang2021topological, krasnok2021parity, bergholtz2021exceptional, parto2021non, ding2022non}, distinct from the more commonly-known ``diabolic point'' degeneracies of Hermitian systems.  At a diabolic point, two eigenstates are orthogonal despite having equal eigenvalues, due to the spectral theorem.  At an EP, however, the eigenstates coalesce and become linearly dependent \cite{heiss2012physics}, as seen in Eq.~\eqref{H2by2evec}; $\mathcal{H}$ is then a ``defective matrix''.

\textbf{Case II}: $\gamma=0$, with $\delta \ne 0$.  Here, the non-Hermiticity originates from nonreciprocal couplings: to hop from the first mode to the second mode, the transition amplitude has a different magnitude than for the reverse.  (In the Hermitian case, the two couplings must have the same magnitude, even if their phases differ.)

Unlike Case I, the Hamiltonian satisfies $[\hat{T}, \mathcal{H}]= 0$, which implies that if $E_+$ is an eigenvalue with right eigenstate $|\psi_+\rangle$, then $E_- = E_+^*$ is another eigenvalue with right eigenstate $\hat{T}|\psi_+\rangle$.  Moreover, $\mathcal{H}$ obeys another symmetry called \textit{pseudo-anti-Hermiticity}: $\eta \mathcal{H} \eta^{-1}=-\mathcal{H}^\dagger$ where $\eta=\sigma_2$.  This implies that for each left eigenstate $\langle\psi_\pm^L|$, we have $\eta^{-1}|\psi_{\pm}^L\rangle$ as a right eigenstate with eigenvalue $-E_\pm^*$ (see Section~\ref{sec:PH}).

This combination of symmetries implies that both eigenvalues are either real (when $\omega^2+\kappa^2>\delta^2$) or purely imaginary (when $\omega^2+\kappa^2<\delta^2$).  The transition point between the two regimes, $\omega^2+\kappa^2=\delta^2$, is an EP.

\textbf{Case III}: $\delta \ne 0$ and/or $\gamma \ne 0$.  The Hamiltonian is non-Hermitian, without any special symmetry.  Nonetheless, the dependence of the eigenstates with $\gamma$, $\omega$, and/or $\kappa$ is influenced by the presence of EPs, as described in Sec.~\ref{sec:ep2}.

\subsection{Exceptional Points}
\label{sec:ep2}

Consider the Hamiltonian $\mathcal{H}$ of Eq.~\eqref{H2by2}, with $\delta = 0$ and $\gamma \ne 0$.  We can see from Eqs.~\eqref{H2by2eval}--\eqref{H2by2evec} that the eigenstates coalesce (i.e., become linearly dependent) at the point $\omega = 0$, $|\kappa| = |\gamma|$.  As previously stated, this is called an exceptional point (EP) \cite{heiss2012physics}.  This EP is specifically called an ``order-2 EP'', or EP$_2$, since it involves the coalescence of two eigenvalues, along with their eigenstates, as illustrated in Fig.~\ref{fig:EP2}(b).

The present Hamiltonian is $2\times2$, so every EP is an EP$_2$.  Larger Hamiltonians can have higher-order EPs involving three or more eigenstate, as discussed in Sec.~\ref{sec:HEPs}.

EPs are an important and commonly-encountered feature of non-Hermitian Hamiltonians.  If two (or more) bands of eigenvalues/eigenstates meet at an EP somewhere in their parameter space, the bands are no longer distinct mathematical objects, but are connected smoothly to each other.  In the above example, consider the $\omega$-$\kappa$ parameter space, for fixed $\gamma$.  As shown in  Fig.~\ref{fig:EP2}(c), following a $2\pi$ loop in the parameter space encircling the EP$_2$ exchanges the two eigenstates, as well as their eigenvalues.  If the loop does not encircle the EP$_2$, each eigenstate returns to itself \cite{dembowski2001experimental, zhong2018winding}.

This exchange property can be understood to originate from the branch structure of the complex eigenvalue spectrum, based on the complex square root operation $\sqrt{z}$ (which has two branches meeting at the branch point $z = 0$).  Some recent theoretical works have sought to describe it in a more sophisticated way, in terms of the geometrical characteristics of the eigenstates under parametric variation \cite{shen2018topological}.  By drawing analogies from the influential theory of geometrical phases and holonomies in Hermitian physics \cite{cohen2019geomreview}, one can develop a non-Hermitian version of the Berry connection, and hence a set of non-Hermitian Berry phases.  Under a two-cycle encirclement of the EP$_2$, the non-Hermitian Berry phase is $\pi$, whereas if the EP$_2$ is not encircled it is zero \cite{shen2018topological}.  This formalism subsequently aids in the construction of a theory of non-Hermitian band topology (see Section~\ref{sec:topology}).

The exchange of eigenstates/eigenvalues under EP encirclement has been mapped out in numerous experiments \cite{dembowski2001experimental, lee2009observation, gao2015observation, hu2017exceptional, yu2021general, ergoktas2022topological}, beginning with a pioneering experiment by Dembowski \textit{\textit{et al.}}~on a tunable microwave cavity \cite{dembowski2001experimental}.  It has been proposed as a novel ingredient for designing photonic structures, e.g.~guiding the layout of meta-atoms on a non-Hermitian metasurface \cite{song2021plasmonic}.  It is worth noting, however, that \textit{dynamically} encircling an EP$_2$ generally does not induce dynamical mode-switching, but rather asymmetric mode conversion (i.e., preferential conversion to the higher-gain mode), due to the breakdown of adiabaticity in non-Hermitian dynamical systems \cite{uzdin2011observability, milburn2015general, doppler2016dynamically, xu2016topological, ghosh2016ep, hassan2017dynamically, wang2018non, yoon2018time, ergoktas2022topological}.

EPs have numerous other interesting physical consequences, from generating large-amplitude transient oscillations under time evolution \cite{makris2014transient} to enabling ``unidirectional invisibility'' in non-Hermitian gratings \cite{lin2011unidirectional}. One noteworthy feature that may hold considerable technological promise is the square root scaling of energies close to an EP$_2$: $|E_\pm| \propto r^{1/2}$, where $r \rightarrow 0$ is the parametric distance to the EP$_2$ [see Eq.~\eqref{H2by2eval}].  By contrast, diabolic points have linear (or higher) scaling with $r$.  Due to the rapid variation of the square root function at small $r$, positioning an eigenmode near an EP$_2$ can enhance its performance as a sensor \cite{wiersig2014enhancing, wiersig2016sensors, chen2017exceptional, zhong2019sensing, hokmabadi2019non, hokmabadi2019non, lai2019observation, park2020symmetry, qin2021exceptional}.  It should also be noted, however, that the sensitivity of a working sensor also depends on the noise level.  For sensors operating at the fundamental noise floor imposed by quantum shot noise (which stem from unavoidable quantum fluctuations), it has been shown that the EP also enhances the noise level, which can reduce or eliminate the benefit of the rapid $r$-scaling \cite{langbein2018no, mortensen2018fluctuations, lau2018fundamental, wolff2019time, xiao2019enhanced, wiersig2020review, zhang2019quantum, chen2019sensitivity, wang2020petermann}. 

\subsection{Higher-Order Exceptional points}
\label{sec:HEPs}

Non-Hermitian Hamiltonians can host higher-order EPs, denoted by $\textrm{EP}_n$, involving the coalescence of $n > 2$ eigenstates (and their associated eigenvalues) \cite{demange2011signatures, ding2016emergence}.  Take the 3-mode Hamiltonian \cite{hodaei2017}
\begin{equation}
\mathcal{H}=\begin{pmatrix}
m + ig & \kappa & 0 \\ \kappa & 0 & \kappa\\  0 & \kappa & -m-ig \\
\end{pmatrix},
\label{HO-1}
\end{equation}
where $\pm m$ and $\pm g$ are the detuning and gain/loss on the first/third modes, and $\kappa$ is a coupling between neighboring modes. At $m = 0$, $|g|= \sqrt{2}\, |\kappa|$, an $\textrm{EP}_3$ occurs; all three right eigenstates coalesce to $[1,-i\sqrt{2}C,-1]^T/2$ where $C=\textrm{sgn}(g/\kappa)$, with eigenvalue $E = 0$.

As for EPs of order $n > 3$ \cite{xiao2019anisotropic, zhong2020hierarchical, delplace2021symmetry}, a systematic way to generate them is to use the Hamiltonian \cite{wang2019arbitrary}
\begin{equation}
\mathcal{H}=\begin{pmatrix}
0 & \kappa_{1,2} & \cdots & \kappa_{1,n} \\ 0 & 0 &\cdots & \kappa_{2,n}\\  \vdots  & \vdots & \ddots & \vdots \\ 0  & 0 & 0 & 0\\
\end{pmatrix}.
\label{HO-3}
\end{equation}
Note that the couplings are all nonreciprocal.  This hosts an EP$_n$ at $E=0$. For example, taking $\kappa_{j,j+1} = 1$ for $j = 1,\dots, n-1$, with all other $\kappa_{ij} = 0$, Eq.~\eqref{HO-3} becomes a Jordan normal form with zeros on the diagonal and geometric multiplicity 1; such a matrix must be defective, with $n$ denegerate eigenvalues/eigenvectors \cite{graefe2008non}.

Another systematic way to produce a higher-order EP is to impose a nonreciprocal coupling between two EPs of the same order \cite{zhong2020hierarchical, wiersig2022revisiting}.  Consider
\begin{equation}
\mathcal{H}=\begin{pmatrix}
H_a & K \\ 0 & H_b 
\end{pmatrix},
\label{HO-4}
\end{equation}
where $H_a$ and $H_b$ are $N \times N$ matrices that both have order-$N$ EPs at the same eigenvalue $E$, with eigenvectors $\psi_a$ and $\psi_b$ respectively; and $K$ is a coupling matrix.  We look for an unnormalized eigenvector of the form $[c_a, c_b]^T$.  There are two distinct possibilities.  First, if $c_b \ne 0$, applying $\mathcal{H}$ to the eigenvector reveals that $c_b = \psi_b$, the eigenvalue is $E$, and $(E-H_a) c_a = K \psi_b$.  But the latter equation may lack a solution for $c_a$ because $E-H_a$ is defective and hence non-invertible.  If so, we fall back on the second possibility, $c_b = 0$; for this case, Eq.~\eqref{HO-4} implies that there is a single eigenvector $[\psi_a, 0]^T$ with eigenvalue $E$, which is an $\mathrm{EP}_{2N}$.

As noted in Sec.~\ref{sec:ep2}, a key feature of an $\textrm{EP}_2$ is the square root scaling of the eigenvalues with parametric distance $r$.  For higher-order EPs, the scaling is $r^{1/\alpha}$ where $\alpha \le n$, with the bound imposed by the order of the characteristic polynomial of $\mathcal{H}$.  The actual value of $\alpha$ is model-dependent (e.g., for the EP$_{3}$ in Eq.~\eqref{HO-1}, we have $\alpha=2$), and can also depend on the direction in parameter space \cite{ding2018experimental}.  Moreover, encircling a higher-order EP can produce much more complicated behaviors than the EP$_2$ case described in Sec.~\ref{sec:ep2}; for instance, different loop choices can induce different eigenstate exchanges \cite{demange2011signatures, tang2020exceptional}.  Experiments on acoustic resonators have been notably useful for investigating these interesting features of higher-order EPs, as well as methods for systematically splitting and combining EPs \cite{ding2016emergence, tang2020exceptional}.

\section{Non-Hermitian Symmetries}
\label{sec:Symmetry}

In Hermitian systems, symmetries of the Hamiltonian have important implications for the spectrum of energy eigenvalues, as well as their eigenstates.

For example, a Hamiltonian is time-reversal symmetric if it obeys the commutation relation
\begin{equation}
  [\hat{T}, \mathcal{H}] = 0,
  \label{treversal}
\end{equation}
for an antiunitary operator $\hat{T}$ called the time-reversal operator, which satisfies $\hat{T}^2=\pm1$.  If $\hat{T}^2=-1$, which is the case for spin-half particles, Eq.~\eqref{treversal} gives rise to Kramer's theorem: given that $\mathcal{H}$ is Hermitian, every energy eigenstate $|\psi\rangle$ is accompanied by an orthogonal eigenstate $\hat{T}|\psi\rangle$ of the same energy \cite{altland1997nonstandard}.

Another important symmetry is chiral symmetry, which is defined by an anticommutation relation
\begin{equation}
  \{\hat{C}, \mathcal{H}\} = 0,
  \label{chirality}
\end{equation}
where the chiral symmetry operator $\hat{C}$ is unitary and satisfies $\hat{C}^2=1$.  If a system obeys chiral symmetry, then for every energy eigenstate $|\psi\rangle$ with energy $E$, Eq.~\eqref{chirality} implies that $\hat{C}|\psi \rangle$ is an eigenstate with energy $-E$.  Hence, the energy spectrum is symmetric around $E = 0$.  This also raises the possibility of self-symmetric states obeying $\hat{C}|\psi\rangle = \pm |\psi\rangle$, for which Eq.~\eqref{chirality} implies that the energy must be pinned to $E = 0$.  Chiral symmetry commonly arises in lattices containing a bipartite symmetry,  with Hamiltonians of the form
\begin{equation}
  \mathcal{H} = \begin{pmatrix}0 & K_1 \\ K_2 & 0 \end{pmatrix}.
\end{equation}
Here, the sub-Hamiltonian governing each of the two sublattices is zero, and $K_1$ and $K_2$ are coupling matrices describing arbitrary connections between the two sublattices.  Such Hamiltonians obey Eq.~\eqref{chirality} with $\hat{C} = \sigma_3\otimes\hat{I}$.


Symmetries play a crucial role in the theory of bandstructure topology (see Sec.~\ref{sec:topology}).  In the Hermitian regime, the set of possible topological phases for a lattice is determined by its spatial dimension, and whether it obeys time-reversal symmetry, chiral symmetry, and another symmetry called particle-hole symmetry.  This categorization scheme is commonly referred to as the ``tenfold way'' \cite{altland1997nonstandard, hasan2010colloquium, BansilReview2016}.

Non-Hermitian systems, on the other hand, have access to certain symmetries that do not occur in the Hermitian context.  We have encountered some of these ``non-Hermitian symmetries'' in Sec.~\ref{sec:model}, such as parity/time-reversal (PT) symmetry, the first non-Hermitian symmetry to be extensively studied \cite{bender1998real, christodoulides2018parity}.  Below, we present some of the more prominent non-Hermitian symmetries and their implications.  We will discuss specific non-Hermitian lattices that exhibit these symmetries in Sec.~\ref{sec:Lattices1D}--\ref{sec:Lattices2D}.


\subsection{Parity/Time-Reversal Symmetry}
\label{sec:pt}

A Hamiltonian $\mathcal{H}$ is parity/time-reversal (PT) symmetric if there is a unitary operator $\hat{P}$ and antiunitary operator $\hat{T}$ such that \cite{bender1998real, christodoulides2018parity}
\begin{align}
  [\hat{P}\hat{T}, \mathcal{H}] &= 0, \label{PT-0}\\
  (\hat{P}\hat{T})^2 &= 1.  \label{PT-twofold}
\end{align}
Eq.~\eqref{PT-0} implies that if $|\psi\rangle$ is a right eigenstate of $\mathcal{H}$ with eigenvalue $E$, then $\hat{P}\hat{T}|\psi\rangle$ is a right eigenstate with eigenvalue $E^*$.  This implies two distinct possibilities.  First, if $|\psi\rangle \propto \hat{P}\hat{T}|\psi\rangle$, then $|\psi\rangle$ is an eigenstate of $\hat{P}\hat{T}$ and $E$ is real (unbroken PT symmetry).  Otherwise, $|\psi\rangle$ and $\hat{P}\hat{T}|\psi\rangle$ are a pair of distinct right eigenstates whose eigenvalues form the conjugate pair $(E, E^*)$, which map to each other under $\hat{P}\hat{T}$ (broken PT symmetry).  This notion of unbroken/broken PT symmetry \cite{bender2007making} is inspired by the phenomenon of spontaneous symmetry breaking in Hermitian physics \cite{abud1983geometry, beekman2019introduction}, whereby a many-body ground state violates a symmetry of its Hamiltonian (e.g., a crystalline solid violates continuous translational symmetry).  However, this is an inexact analogy: PT symmetry breaking is not a many-body quantum effect, and is not limited to the ground state (which is anyway poorly-defined for non-Hermitian systems).

The representations of $\hat{P}$ and $\hat{T}$ are system-specific, though in most cases $\hat{T}$ is the complex conjugation operator.  In fact, the features of PT symmetry discussed in the previous paragraph do not require $\hat{P}$ and $\hat{T}$ to correspond physically to parity and time-reversal.  For example, a parity operator should satisfy $\hat{P}^2 = 1$, but consider the unitary operator
\begin{equation}
  \hat{U} = \begin{pmatrix}e^{i\theta} & 0 \\ 0 & -e^{-i\theta}
  \end{pmatrix},
\end{equation}
where $\theta$ is not a multiple of $\pi$.  Evidently, $\hat{U}^2 \ne 1$, but also $(\hat{U}\hat{T})^2 = 1$ where $\hat{T}$ is complex conjugation \cite{Qi2018}.  Take the Hamiltonian
\begin{equation}
  \mathcal{H} = \begin{pmatrix}m+\Delta & it e^{i\theta} \\ it'e^{-i\theta} & m-\Delta \end{pmatrix},
\end{equation}
where $m, \Delta, t, t' \in \mathbb{R}$.  This is non-Hermitian if $t \ne -t'$, and satisfies $[\hat{U}\hat{T}, \mathcal{H}] = 0$.  Accordingly, its eigenvalues $E_\pm = m \pm \sqrt{\Delta^2 -tt'}$ are either real or a conjugate pair.

The concept of PT symmetry can be further generalized by replacing the right hand side of Eq.~\eqref{PT-twofold} with a constant other than unity.  Doing so preserves the logic of the discussion following Eqs.~\eqref{PT-0}--\eqref{PT-twofold} involving ``PT unbroken'' and ``PT broken'' phases.  The case of $(\hat{P}\hat{T})^2=-1$ has an interesting property: in the PT broken scenario, the partner eigenstates $|\psi\rangle$ and $\hat{P}\hat{T}|\psi\rangle$ are orthogonal, i.e.~$\langle\psi|\hat{P}\hat{T}|\psi\rangle = 0$ \cite{brody2013biorthogonal, kawabata2019symmetry}.

\subsection{Anti-PT Symmetry}
\label{sec:apt}

A Hamiltonian $\mathcal{H}$ is anti-PT symmetric if it obeys the anticommutation relation \cite{ge2017symmetry, zhang2019dynamically, kawabata2019topological, zhang2020synthetic, bergman2021observation, wu2021topology}
\begin{equation}
  \{\hat{P}\hat{T}, \mathcal{H}\} = 0,
  \label{APT-0}
\end{equation} 
where $\hat{P}$ and $\hat{T}$ are once again unitary and antiunitary operators satisfying $(\hat{P}\hat{T})^2 = 1$.  Note that this differs from the chiral symmetry relation \eqref{chirality} as $\hat{P}\hat{T}$ is not unitary.  By reasoning in a manner similar to Sec.~\ref{sec:pt}, the eigenvalues can be shown to be either purely imaginary, or forming pairs $(E, -E^*)$ with the eigenstates mapping to each other under $\hat{P}\hat{T}$.

A notable example of anti-PT symmetry arises in bipartite systems whose Hamiltonians have the form \cite{ge2017symmetry}
\begin{equation}
  \mathcal{H}=\begin{pmatrix} iD_1 & V_1 \\ V_2 & iD_2 \end{pmatrix},
  \label{EPT-1}
\end{equation}
where $V_{1,2}$ and $D_{1,2}$ real $N\times N$ matrices, and $D_{1,2}$ are diagonal. Such a Hamiltonian describes two subsystems of $N$ modes each, with each mode subject to only gain or loss (i.e., zero real frequency detuning), and coupling only to the other subsystem.  It satisfies Eq.~\eqref{APT-0} with the ``subsystem exchange'' operator $\hat{P}=\sigma_3 \otimes I_N$, where $I_N$ is the $N\times N$ identity matrix and $\hat{T}$ is the complex conjugation operator.  In this context, the symmetry has been referred to as ``non-Hermitian particle-hole symmetry'', by analogy with the particle-hole symmetry of Hermitian Hamiltonians (which is likewise expressed using an anticommutation relation) \cite{ge2017symmetry}.

Remarkably, the Hamiltonian \eqref{EPT-1} retains its anti-PT symmetry under arbitrary variations in the imaginary diagonal entries of $D_{1,2}$ (i.e., modulations of the local gain/loss) \cite{ge2017symmetry, jeon2020non, ZhuAnomalous2022}.  This holds even in the nonlinear regime, where the diagonal entries are imaginary functions of the state vector.  Physically relevant nonlinearities of this type include gain saturation, so anti-PT symmetric lattices could be useful for controlling laser modes \cite{ZhuAnomalous2022}.

PT and anti-PT symmetry can be incorporated into a more general symmetry of the form \cite{kawabata2019topological}
\begin{equation}
  \hat{P}\hat{T}\mathcal{H}=e^{i\phi}\mathcal{H}\hat{P}\hat{T},
\label{APT-2}
\end{equation}
where $\phi \in [0, 2\pi)$.  The $\phi=0$ case reduces to PT symmetry, and $\phi = \pi$ reduces to anti-PT symmetry.  Eq.~\eqref{APT-2} has been called ``anyonic PT symmetry'' \cite{longhi2019anyonic, gao2019parity, arwas2022anyonic}, based on a loose analogy with anyonic exchange symmetry in many-body quantum mechanics \cite{Wilczek1982}.  It has been realized in systems of coupled resonators, with the $\phi$ parameter implemented using a combination of dispersive and dissipative inter-resonator couplings \cite{gao2019parity, arwas2022anyonic}.  This concept provides a systematic way to tune a system's eigenvalues in the complex plane; by varying $\phi$, the eigenvalues can, for example, be rotated between purely real values (unbroken PT symmetry) to imaginary values (unbroken anti-PT symmetry).

\subsection{Pseudo-Hermiticity}
\label{sec:PH}

Pseudo-Hermiticity is a symmetry that encompasses Hermiticity and PT symmetry, and, crudely speaking, characterizes a Hamiltonian's capacity to produce real eigenvalues \cite{Pauli1943, mostafazadeh2002P1, mostafazadeh2002P2, mostafazadeh2004physical}.  A pseudo-Hermitian Hamiltonian $\mathcal{H}$ is one that satisfies
\begin{equation}
  \eta \mathcal{H}\eta^{-1}=\mathcal{H}^\dagger
  \label{PH}
\end{equation}
for some invertible Hermitian linear operator $\eta$.  Evidently, for $\eta = 1$ this reduces to the definition of Hermiticity.  Moreover, it can be shown that PT symmetry implies pseudo-Hermiticity for some choice of $\eta$ \cite{mostafazadeh2002P1, mostafazadeh2002P2, ZhangJMP2020}.

Much like PT symmetry, pseudo-Hermiticity implies that the eigenvalues of $\mathcal{H}$ are real or occur in conjugate pairs.  To see this, let $|\psi_n^R\rangle$ denote each right eigenstate of $\mathcal{H}$, with eigenvalue $E_n$, and let $\langle\psi_n^L|$ denote the corresponding left eigenstate.  Eq.~\eqref{PH} implies that $\eta^{-1}|\psi_{n}^L\rangle$ is a right eigenstate of $\mathcal{H}$ with eigenvalue $E_n^*$.  Thus, for a non-degenerate $E_n$, either (i) $E_n$ is real and $\eta^{-1}|\psi_n^L\rangle \propto |\psi_n^R\rangle$, or (ii) $E_n$ is not real, in which case $|\psi_{n'} \rangle = \eta^{-1}|\psi_n^L\rangle$ is a distinct right eigenstate with eigenvalue $E_n^*$.  Eq.~\eqref{PH} also implies that
\begin{align}
  \langle \psi_m^R|\eta \mathcal{H}|\psi_n^R\rangle =
  \langle \psi_m^R|\mathcal{H}^\dagger \eta|\psi_n^R\rangle \\
  \Rightarrow \quad
  (E_n-E_m^*) \langle \psi_m^R|\eta|\psi_n^R\rangle = 0.
  \label{PH2}
\end{align}
Therefore, in case (i), every other right eigenstate $|\psi_m^R\rangle$ satisfies $\langle \psi_m^R|\eta|\psi_n^R\rangle = 0$, a property called ``$\eta$-orthogonality''.  In case (ii), $\eta$-orthogonality applies to all other right eigenstates except the partner $|\psi_{n'}^R\rangle$.

A variant of pseudo-Hermiticity, which has been dubbed ``pseudo-anti-Hermiticity'' \cite{esaki2011edge}, is
\begin{equation}
  \eta \mathcal{H}\eta^{-1}=-\mathcal{H}^\dagger.
  \label{PH-3}
\end{equation}
The eigenvalues of such a Hamiltonian are either purely imaginary, or come in negative conjugate pairs $(E,-E^*)$.  For each right eigenstate $|\psi_n^R\rangle$ with eigenvalue $E_n$, $\eta^{-1}|\psi_n^L\rangle$ is also a right eigenstate with eigenvalue $-E_n^*$.  Any other right eigenstate $|\psi_m^R\rangle$ except the negative conjugate partner of $|\psi_n^R\rangle$ (if any) obeys $\langle \psi_m^R|\eta|\psi_n^R\rangle$.  An interesting application of pseudo-anti-Hermiticity in a lattice is described in Sec.~\ref{sec:lattices}.

\subsection{Combinations of Non-Hermitian Symmetries}
\label{sec:CNHS}

Non-Hermitian symmetries can have interesting interactions with each other, or with conventional symmetries, leading to novel physical effects.  We saw an example of this in Sec.~\ref{sec:model}, where the model Hamiltonian \eqref{H2by2evec} can be made to simultaneously obey time-reversal symmetry and pseudo-anti-Hermiticity (Sec.~\ref{sec:PH}).  This implied the existence of two distinct phases, one with real eigenvalues and the other with imaginary eigenvalues.

Another example of combining non-Hermitian symmetries is seen in the Hamiltonian \cite{xue2020non}
\begin{equation}
\mathcal{H}=\begin{pmatrix} a & b & \lambda_+ & c_+  \\ 
b^* & -a^* & c_+^* & -\lambda_+\\
\lambda_- & c_- & a^* & b \\ 
c_-^* & -\lambda_- & b^* & -a\\
\end{pmatrix}
\label{CNH-1}
\end{equation}  
where $a, b,c_\pm \in \mathbb{C}$ and $\lambda_\pm \in \mathbb{R}$.  Given $\Sigma_\mu=\sigma_1 \otimes \sigma_\mu$ ($\mu=0,1,2,3$) where $\sigma_0$ is the $2\times2$ identity, and the complex conjugation operator $\hat{T}$, we can shown that $\mathcal{H}$ obeys $\{\Sigma_3 \Sigma_1 \hat{T}, \mathcal{H}\} = 0$ (anti-PT symmetry) and $\Sigma_0 \mathcal{H} \Sigma_0= \mathcal{H}^\dagger$ (pseudo-Hermiticity).  The properties discussed in Sec.~\ref{sec:apt} and Sec.~\ref{sec:PH} then imply that the eigenvalues either form a set $\{E,E^*,-E,-E^*\}$ with non-real $E$, or two real pairs $\{E_1,-E_1\}$ and $\{E_2, -E_2\}$.  In the latter case, the eigenstates within each pair are orthogonal (in the usual sense, rather than $\eta$-orthogonality): if $|\psi_{+}\rangle$ is a right eigenstate with eigenvalue $E$, anti-PT symmetry implies that its counterpart is $|\psi_{-}\rangle = \Sigma_1 \Sigma_3 |\psi_{+}\rangle$, with eigenvalue $-E$; and these necessarily obey $\langle \psi_- | \psi_+\rangle = 0$. This can be used to devise non-Hermitian Hamiltonians with diabolic points (such as Dirac or Weyl points, with their distinctive topological properties), rather than EPs \cite{xue2020non}.

\begin{figure}
\centering
\includegraphics[width=0.38\textwidth]{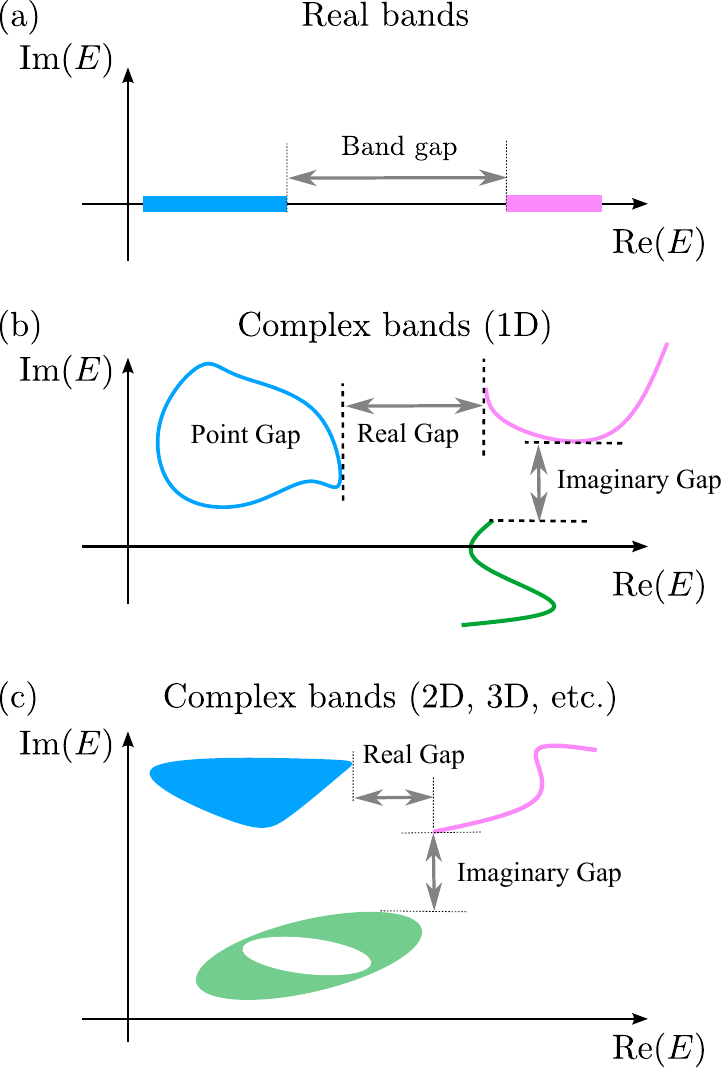}
\caption{Common features of the energy bands of infinite lattices. (a) In Hermitian lattices, and some non-Hermitian lattices such those with unbroken PT symmetry, the eigenenergies all lie on the real line.  A band gap is an energy range between energy bands. (b) In 1D non-Hermitian lattices, bands can form closed loops (blue curve) enclosing point gaps, or open arcs (pink and green curves).  Real gaps and imaginary gaps separate bands in the real and imaginary directions, respectively.  (c) Non-Hermitian lattices in 2D and higher dimensions can have bands filling spectral areas, either with a point gap (green area) or without (blue area), or arcs and loops similar to the 1D case (pink curve).}
\label{fig:Gaptypes}
\end{figure}

\section{Non-Hermitian Lattices}
\label{sec:lattices}

Non-Hermitian lattices are a special class of non-Hermitian systems that obey discrete spatial translational symmetry.  For a spatially infinite lattice, discrete translational symmetry implies that the Hamiltonian can be decomposed as
\begin{equation}
  \mathcal{H} = \sum_{\mathbf{k}} \mathcal{H}_\mathbf{k},
\end{equation}
where $\mathbf{k}$ is a quasimomentum.  The momentum-space Hamiltonians $\{\mathcal{H}_{\mathbf{k}}\}$ define the bandstructure $\{E_n(\mathbf{k}), |\psi_{n\mathbf{k}}\rangle \}$, consisting of eigenenergies and eigenstates labeled by $\mathbf{k}$ and a discrete band index $n$.  If the lattice is Hermitian, then $E_n(\mathbf{k}) \in \mathbb{R}$ and $\langle \psi_{n\mathbf{k}}|\psi_{n'\mathbf{k}}\rangle = \delta_{nn'}$, but this need not hold for the non-Hermitian case.

The quasimomentum index $\mathbf{k}$ in $\mathcal{H}_{\mathbf{k}}$ is significant in a couple of ways.  First, because $\mathbf{k}$ is defined modulo the Brillouin zone, each band is periodic.  For example, a 1D lattice of period $a$ must have $E_n(k) = E_n(k+2\pi/a)$ and $|nk\rangle = |n, k+2\pi/a\rangle$.  Second, by expanding $\mathcal{H}_{\mathbf{k}}$ in $\mathbf{k}$ around certain $\mathbf{k}$ points, the lattice Schr\"odinger equation can be reduced to a partial differential equation that is often physically meaningful and/or tractable.  For example, in the 2D Hermitian bandstructure for graphene, expanding around one of the diabolic points yields a 2D Dirac equation \cite{hasan2010colloquium, BansilReview2016, yan2017topological}.  Both of these features hold for non-Hermitian as well as Hermitian lattices.

Photonics provides some of the most important and relevant platforms for realizing non-Hermitian lattices.  This is due to the relative ease with which complicated photonic structures can be fabricated, as well as the availability of physical processes for introducing controlled amounts of non-Hermiticity (e.g., optical gain media).  These issues will be discussed in Sec.~\ref{sec:platform}.

\subsection{Features of Non-Hermitian Bandstructures}
\label{sec:Gap}

Non-Hermitian bandstructures often possess peculiar features not seen in Hermitian bandstructures.  As noted in Sec.~\ref{sec:ep2}, non-Hermitian bandstructures can (and frequently do) contain EPs, meaning that multiple bands can connect smoothly to one another.  Aside from isolated EPs, bands can also coalesce along exceptional lines or surfaces.  We will see some examples in Sec.~\ref{sec:Lattices1D}--\ref{sec:Lattices2D}.

The band energies of non-Hermitian lattices can be complex, and as a consequence have no natural ordering (absent some symmetry constraint, such as PT symmetry, which can force them onto the real line).
It is often helpful to visualize the complex energy bands by plotting them (i.e., $\{E_{n}(\mathbf{k})\}$ for all $\mathbf{k}$) in the complex energy plane.  As shown in Fig.~\ref{fig:Gaptypes}, they can form one-dimensional curves, called ``spectral arcs'' (for open curves) or ``spectral loops'' (for closed curves), or fill two-dimensional areas, called ``spectral areas'' \cite{zhang2022universal}.  Spectral areas only arise in two- or higher-dimensional lattices, though such systems can also exhibit spectral arcs/loops.

For example, for a PT symmetric lattice whose PT symmetry is unbroken for all $\mathbf{k}$ (see Sec.~\ref{sec:pt}), the band energies are all real and thus form spectral arcs along the real line, as shown in Fig.~\ref{fig:Gaptypes}(a).  As in the Hermitian case, we can identify band gaps as ranges of energy lying between (and not overlapping with) the energy bands.  Similar to the Hermitian case, within this energy range waves cannot propagate, and externally incident waves cannot penetrate the medium \cite{lin2011unidirectional}.

In other situations, the band gap concept must undergo modification \cite{kawabata2019symmetry, bergholtz2021exceptional}.  As shown in Fig.~\ref{fig:Gaptypes}(b)--(c), one can identify a ``real gap'' as an energy range that separates the real parts of the band energies, and likewise an ``imaginary gap'' that separates the imaginary parts.  More generally, one can define a ``line gap'' separating complex energy bands along an arbitrarily-chosen line in the complex plane.

Another generalization of the band gap concept is the ``point gap'', based on identifying a point in the complex plane that is surrounded by (and not overlapping) the complex energy bands.  For instance, a point gap can be enclosed by a spectral loop, as shown in Fig.~\ref{fig:Gaptypes}(b).

Inspired by the theory of topological band insulators \cite{hasan2010colloquium, BansilReview2016}, there has recently been a great deal of theoretical interest in the topological classification of non-Hermitian bands.  Some of these efforts have focused on developing non-Hermitian variants of wavefunction-based band invariants (e.g., Chern numbers), while another approach has focused on the structure of the complex energy bands (e.g., the presence of line gaps and point gaps).  We will discuss these issues in Sec.~\ref{sec:topology}.

Finally, we should note that the focus on spatially infinite lattices and their bandstructures presupposes that these predict the behavior of lattices that are finite, but sufficiently large.  (All experimental samples, after all, are finite.)  This principle holds for Hermitian lattices and a large class of non-Hermitian lattices.  However, it is violated by certain non-Hermitian lattices that exhibit the ``non-Hermitian skin effect'', whereby the energies and eigenfunctions of a finite sample differ qualitatively from those of an infinite lattice.  This phenomenon is discussed in Sec.~\ref{sec:nhse}.

\subsection{1D Non-Hermitian Lattices}
\label{sec:Lattices1D}

\begin{figure}
\centering
\includegraphics[width=0.485\textwidth]{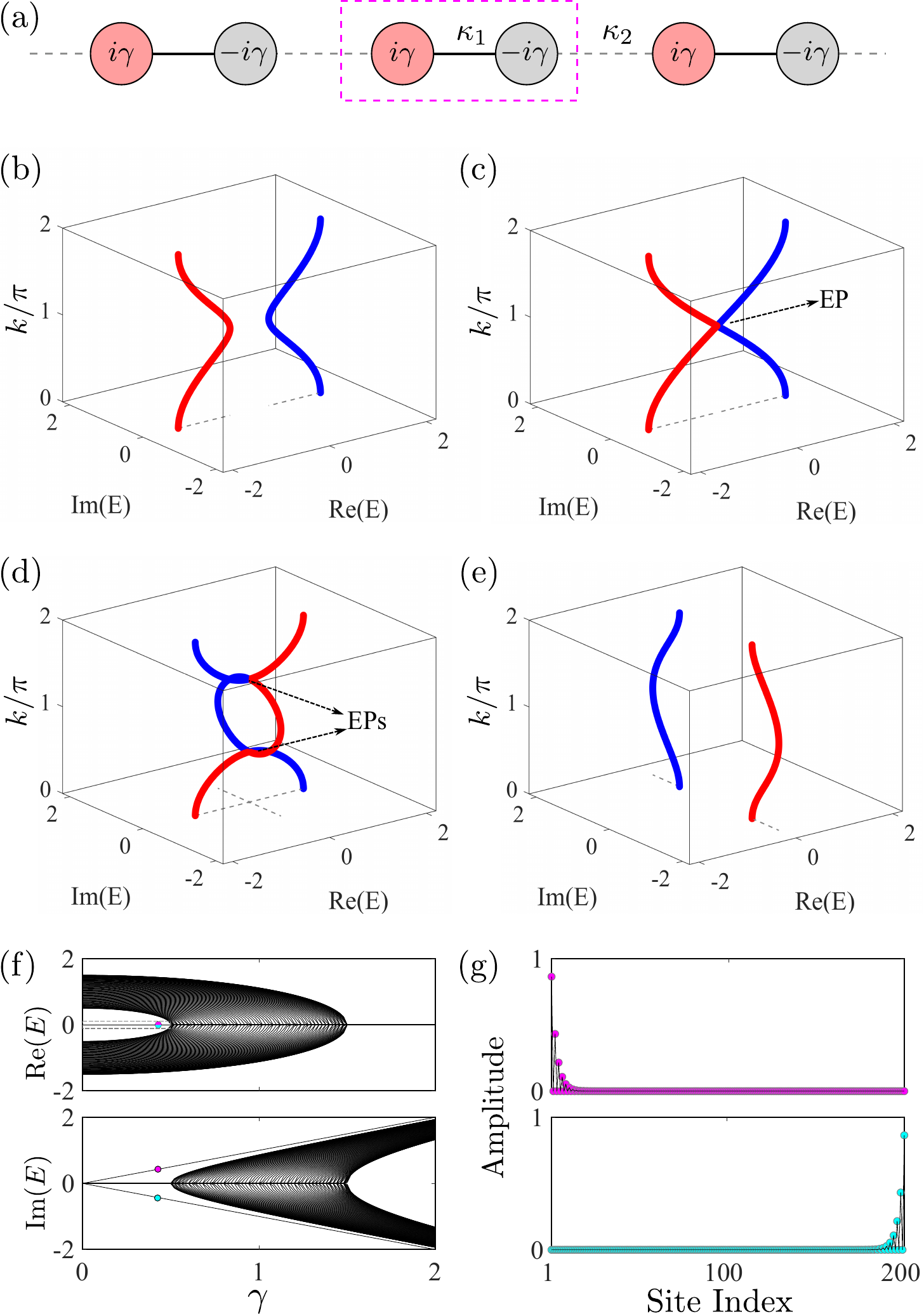}
\caption{(a) Schematic of the complex SSH (cSSH) model with parameter $\kappa_1=0.5$ and $\kappa_2=1$. (b)--(e) Complex energy bands for the cSSH model with (b) $\gamma=0.3$, (c) $\gamma=0.5$, (d) $\gamma=1$, and (e) $\gamma=1.8$.  Dashes indicate the projection of the two bands onto the complex energy plane. (f) Spectrum of the finite cSSH lattice versus $\gamma$,  with real/imaginary parts plotted in upper/bottom panels. (g) Amplitude distributions for two selected edge states with $\gamma=0.4$, corresponding to markers of the same color in (f). The finite lattice in (f) and (g) consists of 200 sites. }
\label{fig:SSHdis}
\end{figure}

Here, and in the following section, we discuss some notable non-Hermitian lattices that have been studied in the literature.  This survey is not meant to be exhaustive; we have selected a small set of models that serve to best illustrate the properties discussed in the rest of this tutorial.  We begin by discussing 1D lattices.

\textit{Complex SSH model}---The Su-Schrieffer-Heeger (SSH) model, originally introduced to describe electronic excitations in polyacetylene \cite{Su1979}, is a workhorse model used in the study of Hermitian topological insulators \cite{hasan2010colloquium, BansilReview2016}.  It consists of a dimer chain formed by alternating strong and weak couplings.  An interesting non-Hermitian extension of this model, called the complex SSH (cSSH) model \cite{schomerus2013topologically, liang2013topological}, involves adding balanced gain/loss to the two sublattices, as shown in Fig.~\ref{fig:SSHdis}(a).

Such a lattice can be implemented, for example, using a 1D array of photonic waveguides with alternating gain and loss \cite{schomerus2013topologically} (see Sec.~\ref{sec:platform}).  Another common alternative is to study a gain-free lattice where one sublattice is lossier than the other, as discussed in Sec.~\ref{sec:waveguide_arrays}.

For the cSSH model with balanced gain/loss, the momentum space Hamiltonian is \cite{schomerus2013topologically, liang2013topological, zhu2014pt, Qi2018}
\begin{equation}
  \mathcal{H}_k=\begin{pmatrix} i\gamma & \kappa_1 +\kappa_2e^{ik}  \\ 
  \kappa_1 +\kappa_2e^{-ik} & -i\gamma \end{pmatrix},
  \label{Example1}
\end{equation} 
where $k \in [-\pi, \pi]$ is the quasimomentum (with the lattice period normalized to unity), $\kappa_1$ and $\kappa_2$ are the real intra- and inter- cell couplings, and $\gamma$ is the magnitude of gain/loss on alternating sites.  This $\mathcal{H}_k$ is PT symmetric (Sec.~\ref{sec:pt}), where $\hat{P}=\sigma_1$ and $\hat{T}$ is the complex conjugation operator.  The PT symmetry is unbroken for all $k$ if and only if
\begin{equation}
  |\gamma| \, <\, \big||\kappa_1|-|\kappa_2|\big|.
\end{equation}
The complex spectrum for this case, which is plotted in Fig.~\ref{fig:SSHdis}(b), has a real gap (Sec.~\ref{sec:Gap}) that is continuable to the gap of the Hermitian ($\gamma = 0$) SSH model with  $\kappa_1=0.5$ and $\kappa_2=1$.  As $|\gamma|$ is increased, EPs appear in the complex energy bands, as shown in Fig.~\ref{fig:SSHdis}(c)--(d) for $\gamma = 0.5$ and $1$ respectively.  For $|\gamma| \, >\, |\kappa_1| + |\kappa_2|$, PT symmetry is broken for all $k$ and the bands become purely imaginary, as shown in Fig.~\ref{fig:SSHdis}(e) with $\gamma = 1.8$.  At this point, the spectrum has an imaginary gap but no real gap. 

In Fig.~\ref{fig:SSHdis}(f)--(g), we plot the eigenvalues of a finite sample of the cSSH lattice with an integer number of unit cells and open boundary conditions (i.e., couplings set to zero at the boundary of the lattice), for
$\kappa_1= 1$, $\kappa_2 = 2$, and $\gamma \in \mathbb{R}^+$.  For $0 < \gamma < \kappa_2 - \kappa_1$, the spectrum has a real gap, corresponding to the aforementioned real gap of the bulk spectrum.  Within this gap, there exist two states with $\mathrm{Re}(E) = 0$; in Fig.~\ref{fig:SSHdis}(f) these are indicated by cyan and purple markers at a specific value of $\gamma$.  We plot the spatial distribution of these midgap states in Fig.~\ref{fig:SSHdis}(g).  They are localized to the boundaries of the lattice, allowing us to deduce that these are non-Hermitian extensions of the topological zero-energy boundary states of the SSH model \cite{schomerus2013topologically}.

The two boundary states map to each other under PT symmetry (see Sec.~\ref{sec:Symmetry}).  Hence, their (imaginary-valued) energies are complex conjugates of each other, as seen in Fig.~\ref{fig:SSHdis}(f).  We can make a crude argument for their existence by considering the $\kappa_1 \rightarrow 0$ limit; from Fig.~\ref{fig:SSHdis}(a), we see that each boundary state will be concentrated on an isolated boundary site, so their energies are $\pm i\gamma$.

To verify the topological nature of the boundary states more rigorously, one can use Eq.~\eqref{Example1} to calculate a non-Hermitian global Berry phase (see Sec.~\ref{sec:wavetopology}).  This yields a value of $2\pi$ in this case; on the other hand, if $\kappa_1 > \kappa_2$, the non-Hermitian global Berry phase is zero and the finite lattice has no boundary states \cite{liang2013topological}.  This result does not depend on $\gamma$, and is thus independent of the PT-breaking transition governed by $\gamma$.

\begin{figure}
\centering
\includegraphics[width=0.48\textwidth]{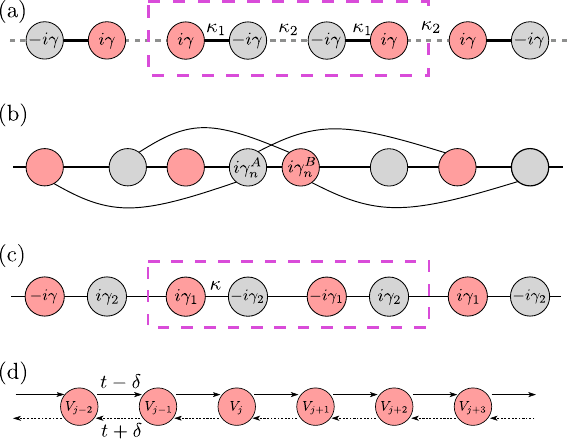}
\caption{Schematics of various 1D non-Hermitian lattices. (a) A non-Hermitian variant of the SSH lattice whose Hamiltonian \eqref{Example2} is anti-PT symmetric \cite{wu2021topology}. The solid and dashed lines indicate the couplings $\kappa_1$ and $\kappa_2$ respectively. (b) A bipartite lattice with purely imaginary on-site potential, whose Hamiltonian Eq.~\eqref{EPT-1} is anti-PT symmetric \cite{LiGe2014prx}. The lines indicate couplings between different sublattices, which need not be limited to nearest-neighbor and can have arbitrary coupling coefficients.  (c) A lattice exhibiting gain/loss-induced topological modes characterized by the non-Hermitian Berry phase \cite{takata2018photonic}. (d) The Hatano-Nelson model, which has nonreciprocal nearest-neighbor couplings (solid and dashed arrows) and exhibits the non-Hermitian skin effect \cite{hatano1996localization, hatano1997vortex}. In all subplots, the color of the circles indicates the sign of the imaginary on-site potential; in (a) and (c), the unit cell is indicated by a dashed magenta box. }
\label{fig:1Dlattice}
\end{figure}

\textit{Anti-PT SSH model}---Another non-Hermitian variant of the SSH model, satisfying anti-PT symmetry (Sec.~\ref{sec:apt}), is depicted in Fig.~\ref{fig:1Dlattice}(a).  It consists of an SSH model overlaid with an alternating pairwise distribution of gain (red circles) and loss (gray circles), resulting in a four-site unit cell
\cite{wu2021topology}.  The momentum space Hamiltonian is
\begin{equation}
\mathcal{H}_k=\begin{pmatrix} i\gamma & \kappa_1 & 0 & \kappa_2e^{ik}  \\ 
\kappa_1 & -i\gamma & \kappa_2 & 0\\
0 & \kappa_2 & -i\gamma & \kappa_1\\
\kappa_2e^{-ik} & 0 & \kappa_1 & i\gamma \end{pmatrix}.
\label{Example2}
\end{equation} 
This anticommutes with $\hat{P}\hat{T}$, where $\hat{P}=i\sigma_1 \otimes \sigma_2$ and $\hat{T}$ is the complex conjugation operator; hence, the spectrum is symmetric around the imaginary axis (see Sec.~\ref{sec:apt}).  Moreover, $\mathcal{H}_k$ is pseudo-anti-Hermitian [see Eq.~\eqref{PH-3} of Sec.~\ref{sec:PH}], with $\eta = \sigma_0 \otimes \sigma_3$.

The model exhibits ``thresholdless'' breaking of the anti-PT symmetry \cite{LiGe2014prx}, meaning that the breaking occurs as soon as $\gamma \neq 0$ (by contrast, the cSSH lattice undergoes PT-breaking at a nonzero value of $\gamma$).  Furthermore, for $\kappa_1 > \kappa_2$ a separate topological transition occurs at $\gamma^2 = \kappa_1^2 - \kappa_2^2$, which corresponds to a real gap closure at $\mathrm{Re}(E) = 0$; when $\gamma$ exceeds this value, the lattice exhibits boundary states with $\mathrm{Re}(E) = 0$.  Unlike the cSSH case, the anti-PT symmetry ensures that the states on opposite boundaries have the same (not conjugate) imaginary energies \cite{wu2021topology}.

Fig.~\ref{fig:1Dlattice}(b) shows another route to anti-PT symmetry which was devised by Ge \cite{ge2017symmetry}.  This design is extremely flexible but generally lacks the topological transitions of the previous model.  It consists of a bipartite lattice, with the sites of each sublattice (drawn as solid and dashed circles in the figure) coupling only to the other sublattice; moreover the on-site potential must be purely imaginary.  With no additional restrictions (e.g., the lattice need not even be periodic), the Hamiltonian has the form of Eq.~\eqref{EPT-1} and is hence anti-PT symmetric.  This scheme is not limited to 1D lattices, and can be applied to higher dimensions \cite{ge2017symmetry}.

\textit{1D Chain with Gain/Loss-Induced Zero Modes}---Fig.~\ref{fig:1Dlattice}(c) depicts a 1D model studied by Takata and Notomi \cite{takata2018photonic}.  Its momentum space Hamiltonian is
\begin{equation}
\mathcal{H}_k=\begin{pmatrix} i\gamma_1 & \kappa & 0 & \kappa e^{-ik}  \\ 
\kappa & -i\gamma_2 & \kappa & 0\\
0 & \kappa & -i\gamma_1 & \kappa\\
\kappa e^{ik} & 0 & \kappa & i\gamma_2 \end{pmatrix}
\label{Example3}
\end{equation} 
This is similar to the model of Eq.~\eqref{Example2}, except that the nearest-neighbor couplings are uniform and the two sets of gain/loss parameters are independently tunable.  Remarkably, the Hamiltonian is both pseudo-Hermitian \textit{and} pseudo-anti-Hermitian (Sec.~\ref{sec:PH}), with different $\eta$ operators \cite{takata2018photonic}.  As a consequence, the eigenvalues are constrained to be purely real pairs $\{E,-E\}$, purely imaginary pairs $\{E,E^*\}$, or a set of the form $\{E, -E, E^*, -E^*\}$.  By tuning to certain nonzero values of $\gamma_1$ and $\gamma_2$, the model can be made to produce topological boundary states (which are characterizable via the non-Hermitian Berry phase).  This is interesting because in the Hermitian limit ($\gamma_1=\gamma_2=0$) the lattice is gapless and has no topological states \cite{takata2018photonic}.

\textit{Hatano-Nelson model}---In the 1990s, Hatano and Nelson studied the properties of quantum systems with imaginary vector potentials \cite{hatano1996localization, hatano1997vortex}.  They developed the influential non-Hermitian 1D model depicted in Fig.~\ref{fig:1Dlattice}(d), which has one site per unit cell and nonreciprocal (i.e., asymmetric) inter-site couplings.  The Hatano-Nelson model is the simplest model that exhibits what is now called the non-Hermitian skin effect, which will be discussed in detail in Sec.~\ref{sec:nhse}.


\begin{figure*}
\centering
\includegraphics[width=0.9\textwidth]{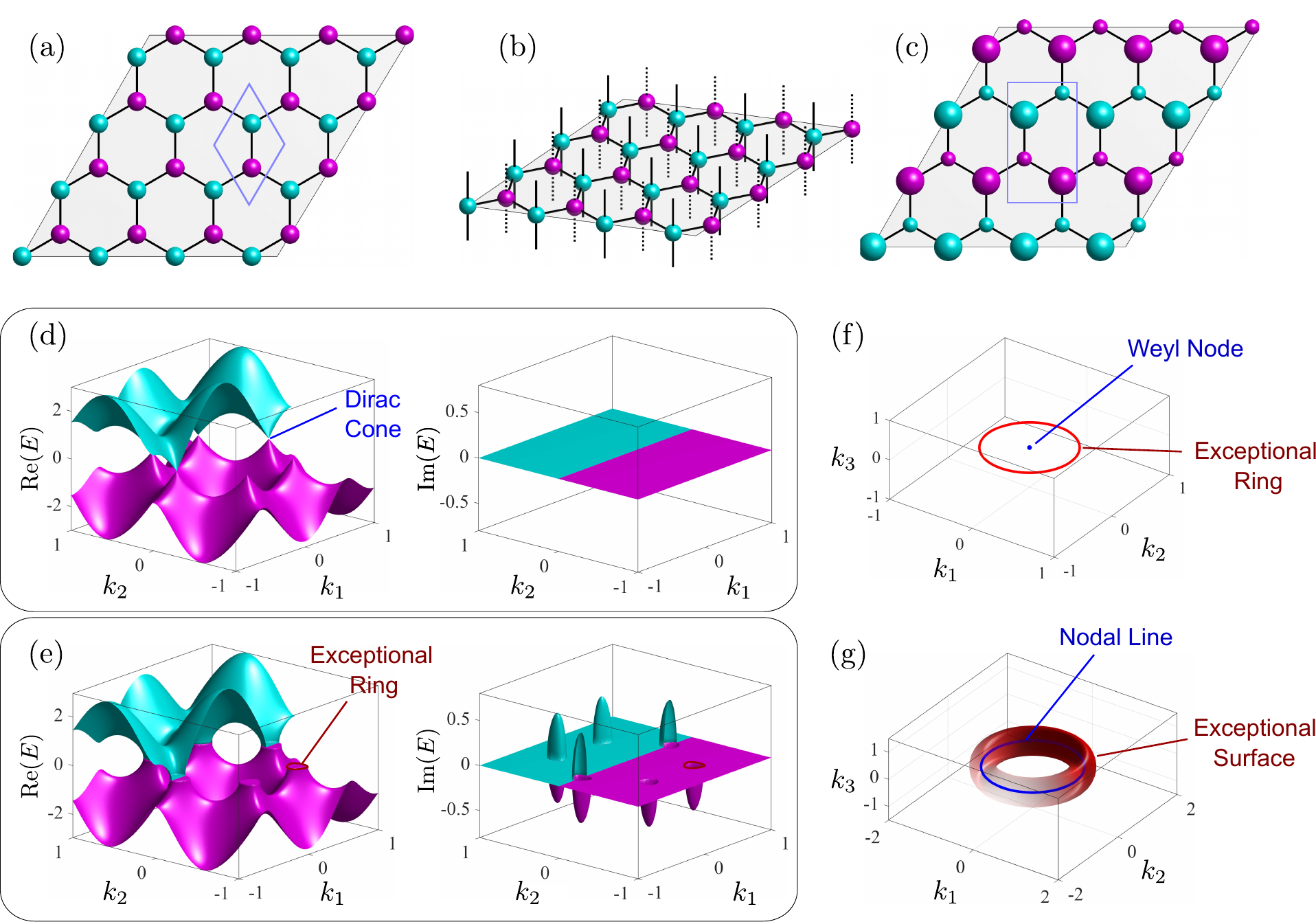}
\caption{Non-Hermitian 2D and 3D lattices and their energy bands. (a) A 2D honeycomb lattice, with each unit cell (blue box) consisting of two sites with different imaginary on-site potentials (cyan and magenta spheres). (b) A 3D lattice formed by stacking the honeycomb lattice of (a) with inter-layer couplings $t_{A/B} \in \mathbb{R}$, as indicated by the solid and dashed lines. (c) A modified 2D honeycomb lattice with each unit cell consisting of four sites, with the color and size of each sphere indicating the imaginary and real part of the on-site potential. (d)--(e) Spectrum of the honeycomb lattice shown in (a) for (d) $\gamma=0$ and (e) $\gamma = 0.5$. The real and imaginary parts are shown in the left and right panels respectively, and the second band is partially removed to show the details.  The nearest-neighbor coupling is set to unity. (f) An exceptional ring formed by Eq.~\eqref{H-ER}, with $\gamma=0.5$.  The blue point indicates the zero-energy Weyl node in the Hermitian ($\gamma = 0$) system.  (g) An exceptional surface formed by Eq.~\eqref{H-NL}, for $\gamma=0.5$.  The surface is a torus enclosing the Hermitian system's zero-energy nodal line (blue loop).}
\label{fig:ERS}
\end{figure*}

\subsection{Higher-Dimensional Non-Hermitian Lattices}
\label{sec:Lattices2D}

Non-Hermitian lattices with two or higher spatial dimensions can exhibit features not found in 1D lattices.  For instance, their energy bands can take on more complicated forms.

Take the honeycomb lattice with uniform nearest-neighbor couplings, depicted in Fig.~\ref{fig:ERS}(a).  In the Hermitian limit, this reduces to a graphene-type semimetal, with Dirac cones at the two inequivalent corners of the Brillouin zone (the $K$ and $K'$ points) \cite{haldane1988model, geim2007rise}, as shown in Fig.~\ref{fig:ERS}(d).  If we introduce gain and loss to the two sublattices, as indicated by the magenta and cyan circles, the 2D lattice is PT symmetric.  Each Dirac cone turns into an exceptional ring---a ring of EPs that encloses the $K$ or $K'$ point \cite{szameit2011p, schomerus2013parity}, as shown in Fig.~\ref{fig:ERS}(e).

Now suppose we stack layers of the honeycomb lattice along the $z$ axis to form a 3D lattice, as shown in Fig.~\ref{fig:ERS}(b). The solid and dashed lines indicate the inter-layer couplings for the two sublattices; if these are unequal, and gain/loss is absent, the 3D Hermitian lattice hosts Weyl points \cite{xiao2015synthetic}, which are points in momentum space around which the Hamiltonian takes the form
\begin{equation}
\mathcal{H}_\mathbf{k} \approx \sum_{i=1}^3 k_i \sigma_i.
\label{H-Weyl}
\end{equation}
Here, $\sigma_i$ are the Pauli matrices, and $\mathbf{k}$ is relative to the Weyl point.  The eigenvalues are $E_{\pm}(\mathbf{k}) =\pm \sqrt{\sum_i k_i^2}$.  The Weyl point at $\mathbf{k} = 0$ is a two-fold band degeneracy \cite{lu2015experimental, noh2017experimental, yan2017topological}.  Upon introducing gain and loss $\pm i\gamma$ into the two sublattice, the $k_3$ term in Eq.~\eqref{H-Weyl} is replaced by $k_3+i\gamma$, and the eigenvalue degeneracy condition becomes
\begin{equation}
  k_1^2 + k_2^2 + (k_3 + i \gamma)^2 = 0.
   \label{H-ER}
\end{equation}
As shown in Fig.~\ref{fig:ERS}(f), this describes a ring in the 3D momentum space, $k_1^2+k_2^2 = \gamma^2$ and $k_3 = 0$, called an exceptional ring (ER) \cite{xu2017, cerjan2018, cerjan2019, kawabata2019classification, mc2020weyl,liu2021, ghorashi2021non, xu2022}.  Although the lattice in this example is PT symmetric, Weyl points in other lattices also typically turn into ERs when subjected to non-Hermitian perturbations \cite{cerjan2018}.

Higher-dimensional generalizations of EPs and ERs, called exceptional surfaces (ESs), can also exist.  Take the Hamiltonian
\begin{equation}
\mathcal{H}_\mathbf{k}=(k_1^2+k_2^2 - m) \sigma_1 + k_3 \sigma_2 + i \gamma \sigma_3,
\label{H-NL}
\end{equation}
whose eigenvalues are
\begin{equation}
  E_\pm=\pm \sqrt{(k_1^2+k_2^2-m)^2+k_3^2-\gamma^2}.
 \label{H-ES} 
\end{equation}
For $\mathbf{k}$ real, $m \in \mathbb{R}+$, and $\gamma = 0$, $\mathcal{H}_\mathbf{k}$ is Hermitian and describes a ``nodal ring semimetal'' \cite{fang2016topological, gao2018experimental}, with degenerate eigenvalues $E_\pm = 0$ along the nodal ring $k_1^2+k_2^2=m$, $k_3=0$.  If we now set $\gamma \ne 0$, then $\mathcal{H}_\mathbf{k}$ becomes non-Hermitian, and its eigenstates (and eigenvalues) coalesce along the ES $(k_1^2+k_2^2-m)^2+k_3^2=\gamma^2$, which is a torus enclosing the initial nodal ring, as shown in Fig.~\ref{fig:ERS}(f).  It is also worth noting that $\mathcal{H}_\mathbf{k}$ is anti-PT symmetric, where $\hat{P}=\sigma_3$ and $\hat{T}$ is complex conjugation (see Sec.~\ref{sec:apt}).  If we add a term $i\gamma \sigma_2$ to the Hamiltonian, the anti-PT symmetry is spoiled and the ES breaks up into an ER.

There are numerous other models that can exhibit exceptional lines, rings, and surfaces \cite{carlstrom2018exceptional, carlstrom2018exceptional, kawabata2019classification, carlstrom2019knotted, budich2019symmetry, rui2019pt, yang2019non, wang2019non, zhou2019,  okugawa2019topological, wang2020exceptional,  yang2021scalable, wu2022non, cui2022symmetry}.  For example, Rui \textit{\textit{et al.}}~discussed a four-band non-Hermitian Dirac Hamiltonian that hosts an ER or ES while also preserving PT symmetry \cite{rui2019pt}.  Exceptional curves in 3D momentum space can form links and and knots \cite{carlstrom2018exceptional, yang2019non, carlstrom2019knotted, hu2021knots}.  There have also been many experimental realizations of these phenomena \cite{zhen2015spawning, cerjan2019, wang2021simulating, zhang2019experimental, zhong2019sensing, qin2021exceptional}, such as using lossy photonic crystal slabs \cite{zhen2015spawning} and optical waveguide arrays \cite{cerjan2019}.

In Sec.~\ref{sec:CNHS}, we discussed a combination of non-Hermitian symmetries (anti-PT symmetry and pseudo-Hermiticity) that counter-intuitively allows for the formation of diabolic points, rather than the EP/ER/ES degeneracies that non-Hermitian systems typically sport \cite{xue2020non}.  This can be achieved using the modified honeycomb lattice shown in Fig.~\ref{fig:ERS}(c). Each unit cell (marked by the blue rectangle) has four sites, with on-site potentials $\{m+i\gamma, -m+i\gamma, m-i\gamma, -m-i\gamma \}$.  The resulting bandstructure can host Dirac points, and opening a gap (which can be done, for instance, by varying the gain/loss) yields conventional or topological insulator phases much like those of Hermitian lattices.  Similarly, stacking layers of this lattice along $z$, similar to Fig.~\ref{fig:ERS}(b), produces a 3D lattice with Weyl nodes sustained by the combined non-Hermitian symmetries \cite{xue2020non}.

\section{Non-Hermitian Band Topology}
\label{sec:topology}

\subsection{Non-Hermitian Wavefunction Topology}
\label{sec:wavetopology}

The theory of band topology, which originated in condensed matter physics, classifies materials into discrete ``topological phases'' according to whether their bandstructures can be smoothly deformed into each other \cite{hasan2010colloquium, BansilReview2016}.  It explains many important phenomena, such as the robust quantization of the Hall conductance in the integer quantum hall effect \cite{klitzing1980new}, which is tied to nonzero integer topological invariants (Chern numbers) of 2D electronic bands.  The physical significance of band topology is summarized by the ``bulk-boundary correspondence principle'', which states that given two materials with a shared bandgap but topologically distinct bandstructures, there must exist in-gap ``topological states'' localized to the interface between the materials.  The band topology concept has been fruitfully applied to photonics \cite{lu2014topological, ozawa2019topological}, as well as other classical wave platforms (e.g., acoustics \cite{ma2019topological, xue2022topological} and electric circuits \cite{imhof2018topolectrical, wang2020circuit}).

The standard theories of band topology assume Hermiticity, but in some cases they remain useful in the non-Hermitian regime.  For example, 1D lattice models like the SSH model \cite{Su1979} (see Sec.~\ref{sec:Lattices1D}) can be characterized using the Berry phase or Zak phase, which is the integral of the Berry connection
\begin{equation}
  A_{k} = i\langle\psi_{nk} |\partial_{k}| \psi_{nk} \rangle
  \label{berryconn}
\end{equation}
across the 1D Brillouin zone.  The Zak phase is quantized to $0$ or $\pi$ (modulo $2\pi$), and a nonzero value implies that a finite (but sufficiently large) sample of the lattice must exhibit topological boundary states with energies lying in the bulk bandgap \cite{hasan2010colloquium, BansilReview2016}.

Rudner and Levitov found that by adding loss to certain sites on a 1D lattice, the topological phase can be made to influence the time-evolution of a wavefunction within the lattice \cite{rudner2009topological}.  This established a novel experimental signature for the Zak phase, one independent of boundary effects.  This prediction was subsequently verified experimentally using a lossy optical waveguide array \cite{zeuner2015observation} (see Sec.~\ref{sec:waveguide_arrays}).  Using similar ideas, the signatures for other Hermitian topological invariants, such as the Chern number, might also be detectable in lossy photonic lattices \cite{leykam2021probing}. 

From a theoretical point of view, however, Eq.~\eqref{berryconn} can be seen to be ill-suited for describing general non-Hermitian lattices.  When the Hamiltonian $\mathcal{H}_{k}$ is Hermitian, the $\langle\psi_{nk}|$ in Eq.~\eqref{berryconn} is both the conjugate of the eigenstate $|\psi_{nk}\rangle$ and a left eigenstate of $\mathcal{H}_{k}$.  If $\mathcal{H}_{k}$ is non-Hermitian, however, Eq.~\eqref{berryconn} is ambiguous since the left and right eigenvectors are generally distinct (see Sec.~\ref{sec:model}).  We can define the non-Hermitian Berry connections
\begin{equation}
  A_{nk}^{\alpha \beta} = i\langle \psi_{nk}^\alpha|\partial_k |\psi_{nk}^\beta \rangle,
  \label{NHberryconn}
\end{equation}
where $\alpha, \beta \in \{L, R\}$ indicate right or left eigenvectors.  It can be shown that $A_{nk}^{LR}=-(A_{nk}^{RL})^*$.  Hence, we can define the non-Hermitian Berry phases
\begin{equation}
  \gamma_n^{\alpha \beta}= \int dk \;A_{nk}^{\alpha \beta},
\end{equation}
where the integral is carried out over the 1D Brillouin zone.  In many models, such as the cSSH model discussed in Sec.~\ref{sec:Lattices1D}, it is found that the ``global Berry phase''
\begin{equation}
  Q=\frac{1}{2\pi}(\gamma_1^{LR}+\gamma_2^{LR})
\end{equation}
is quantized to $0$ or $1$, with a nonzero value predicting the existence of boundary states \cite{liang2013topological, yin2018geometrical, takata2018photonic}.

Similarly, for 2D lattices the Hermitian Berry connection is used to define the Berry curvature $\nabla_{\mathbf{k}}\times\mathbf{A}_{\mathbf{k}}$ \cite{hasan2010colloquium, BansilReview2016}, and integrating this over the 2D Brillouin zone yields the Chern number (an integer-valued topological invariant).  For an isolated non-Hermitian band, the four differents ways of defining the non-Hermitian Berry curvature, based on different combinations of the left and right eigenstates, turn out to yield the same result for the Chern number \cite{shen2018topological, wu2019inversion}.  An interface between two lattices with different Chern numbers hosts topological edge states, whose energies can be complex-valued.


A more extensive reformulation of band topology is required to describe other kinds of non-Hermitian lattices.  As mentioned in Sec.~\ref{sec:Symmetry}, topological insulators in the Hermitian regime can be categorized based on their spatial dimension, as well as whether they obey time-reversal, chiral, and particle-hole symmetries (the possible combinations of symmetry combinations constitute ten distinct ``Altland-Zirnbauer classes'') \cite{altland1997nonstandard}.  Kawabata \textit{et al.}~have extended this analysis to the non-Hermitian regime, and concluded that the ten Altland-Zirnbauer classes should be expanded into 38 classes, in order to account for the richer set of available non-Hermitian symmetries \cite{kawabata2019symmetry}.  Hence, non-Hermitian systems can support a richer variety of topological insulator phases than Hermitian ones.

In Hermitian lattices, topological crystalline insulators \cite{fu2011topological} are topological insulators that impart topological protection through crystal point group symmetries rather than internal symmetries, and thus fall outside of the Altland-Zirnbauer classification scheme.  Much theoretical effort has recently been devoted to the study of non-Hermitian topological crystalline insulators \cite{luo2019higher, edvardsson2019non, ezawa2019non, liu2019second, yu2021zero, wu2021floquet, xu2022edge}.  For example, Liu \textit{et al.}~found that non-Hermitian high-order topological insulators can exhibit topological states at only one corner, unlike their Hermitian counterparts \cite{liu2019second}.

Another class of topological phases falling outside the Altland-Zirnbauer classification are the so-called topological semimetals, which are ungapped materials with topologically charged band degeneracies \cite{hasan2010colloquium, BansilReview2016}.  This includes the Weyl semimetals described in Sec.~\ref{sec:Lattices2D}, which host topologically nontrivial Weyl points that give rise to Fermi arc surface states.  In the non-Hermitian regime, EPs and other non-Hermitian band degeneracies can also be associated with topological modes, distinct from those found in Hermitian systems.  As an example, consider a 2D non-Hermitian medium governed by the $2\times2$ momentum space Hamiltonian \cite{leykam2017edge}
\begin{equation}
\mathcal{H}_{\mathbf{k}} = (k_x+i s k_y) \sigma_1 + m \sigma_3,
\label{NHD-1}
\end{equation}
where $m$ is real and $s = \pm 1$.  The spectrum is complex and contains two EPs at $k_x = 0$, $k_y = \pm |m|$.  If we take two such media with different $m$ and/or $s$, placing them at $x > 0$ and $x < 0$ respectively, it is possible to find a family of zero modes (solutions with $E = 0$) tied to the EPs.  For instance, if we set $m = m_0 > 0$ on the left and $m = -m_0$ on the right, with $s = 1$ everywhere, the zero mode wavefunctions are
\begin{equation}
  \psi(x,y,k_y) = \begin{pmatrix} i \\ 1 \end{pmatrix}e^{-\kappa(x)\, x+ik_y y}
\label{NHD-4}
\end{equation}
where $\kappa(x) = -\mathrm{sgn}(x) m_0 - k_y$.  This is continuous at $x = 0$, and can be shown using the standard prescription $\mathbf{k} \leftrightarrow -i \nabla$ to be a zero-energy eigenstate of the Hamiltonian \eqref{NHD-1} on both sides.  For $-m_0 < k_y < m_0$, the wavenumber range spanning the projections of the EPs, the zero modes are normalizable and localized to the domain wall.  A more rigorous argument, based on counting normalizable states via an index theorem, similarly shows that the zero modes are closely linked to the EPs \cite{leykam2017edge}.

Zero modes can also be found in 3D non-Hermitian Weyl and nodal line media \cite{gonzalez2017topological, luo2018nodal, ghorashi2021non, liu2021, song2022weyl}.  Non-Hermitian higher-order Weyl media, which host zero-energy ``hinge modes'' bounded by ERs, have also been studied theoretically \cite{liu2021}.

\subsection{Spectral Topology and the Non-Hermitian Skin Effect}
\label{sec:nhse}

Non-Hermitian lattices can exhibit another type of band topology, associated with the topology of the complex energy spectrum \cite{yin2018geometrical, zhang2020correspondence, borgnia2020non, okuma2020topological, wojcik2021eigenvalue, zhang2022universal}.  As discussed in Sec.~\ref{sec:Gap}, non-Hermitian energy bands can have different shapes in the complex plane, such as arcs, loops or areas.  These can be categorized into different topological classes, resulting in a topological classification distinct from the wavefunction-based topology described in Sec.~\ref{sec:wavetopology}.  Spectral topology is unique to non-Hermitian systems because of their ability to support complex-valued eigenenergies.

\begin{figure}
\centering
\includegraphics[width=0.48\textwidth]{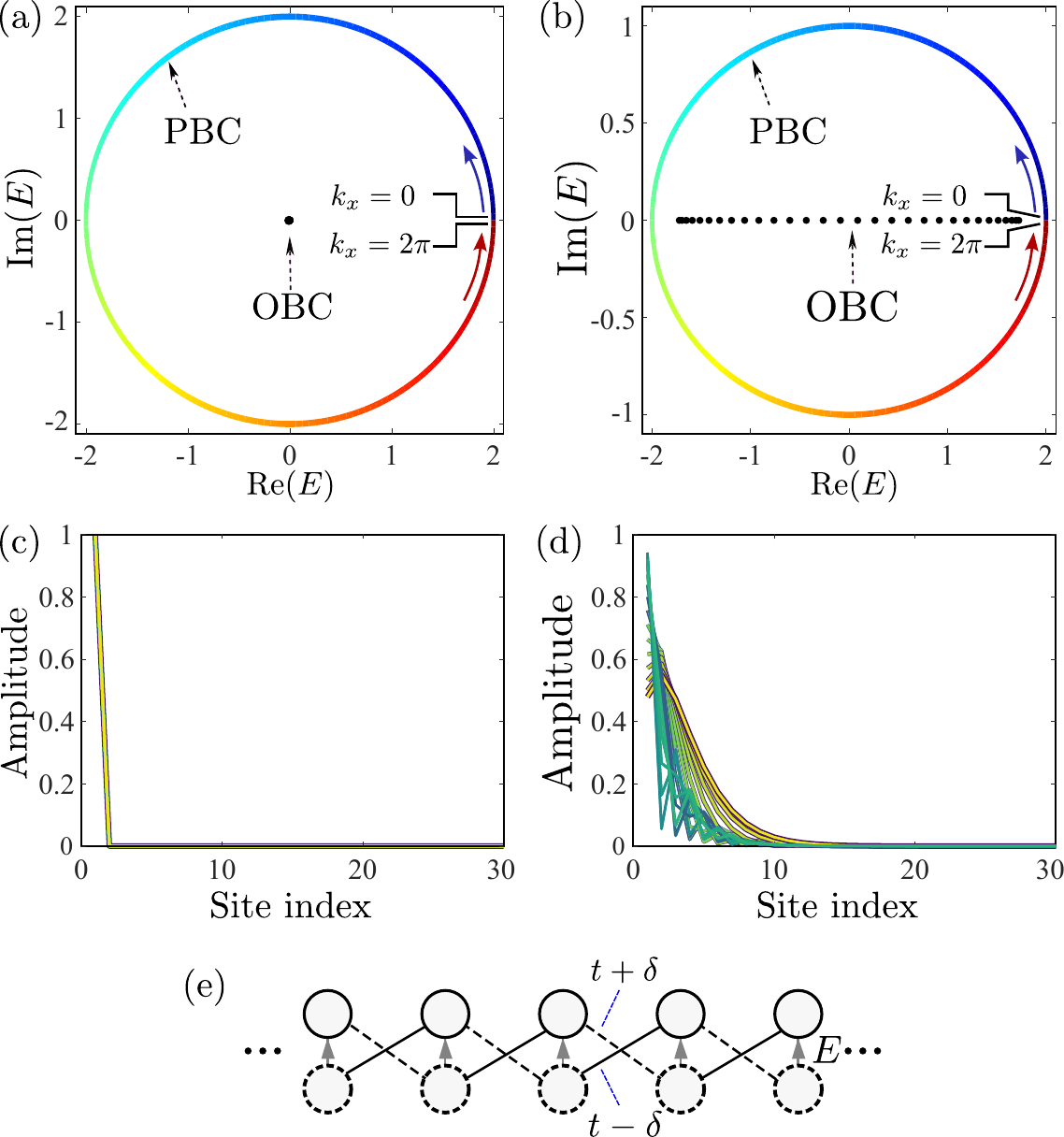}
\caption{(a)--(b) Complex energy spectrum of the Hatano-Nelson model for $t = 1$ and (a) $\delta=1$ and (b) $\delta=0.5$.  Colored curves show the infinite lattice (PBC) spectrum with colors indicating the value of $k \in [0, 2\pi)$.  Black dots show the eigenenergies for a finite lattice of 30 sites, with OBC.  (c)--(d) Spatial distribution of the eigenstate amplitudes for the finite lattices of (a)--(b).  Each eigenstate is plotted in a different color.  (e)  Schematic of the Hermitian ladder lattice defined by Eq.~\eqref{HN-2}, whose zero modes map to eigenstates of the Hatano-Nelson model.  }
\label{fig:HNmodel}
\end{figure}

In 1D models, spectral topology is closely associated with a phenomenon known as the non-Hermitian skin effect (NHSE), whereby an extensive number of bulk eigenstates become localized to the lattice boundary \cite{yao2018edge, jin2019bulk, jiang2019interplay, longhi2019probing, imura2019generalized, lee2019hybrid,  song2020two, zhu2020photonic, li2020critical, li2020topological, mandal2020nonreciprocal, okugawa2021non, xiao2022topology, longhi2022self, xue2022non, lv2022curving}.  We will discuss the 1D NHSE in the rest of this subsection.  In 2D or higher dimensions, the relationship between the NHSE and spectral topology remains unsettled \cite{zhang2022universal}, as we will discuss in Sec.~\ref{sec:nhse2d}.

The paradigmatic model exhibiting the NHSE is the Hatano-Nelson model \cite{hatano1996localization, hatano1997vortex}, which was previously mentioned in Sec.~\ref{sec:Lattices1D} and depicted in Fig.~\ref{fig:1Dlattice}(d).  The model has a single site per unit cell, and the nonzero elements in the Hamiltonian are
\begin{align}
  \mathcal{H}_{j\pm1,j} &= t \pm \delta,\\
  \mathcal{H}_{jj} &= V_j,
  \label{HN-1}
\end{align}
where $j$ is the site index and $V_j, t,\delta \in \mathbb{R}$.  For $\delta \ne 0$, the nearest-neighbor couplings have unequal magnitudes: $|\mathcal{H}_{j+1,j}| \ne |\mathcal{H}_{j,j+1}|$.  Such couplings are ``nonreciprocal'' as they violate the principle of reciprocity in wave propagation \cite{fan2012reciprocity}.  Note that non-Hermitian systems are not necessarily nonreciprocal, and nonreciprocity similarly does not imply non-Hermiticity; in this case, however, the $\delta \ne 0$ model is both nonreciprocal and non-Hermitian.

The Hamiltonian \eqref{HN-1} can be interpreted as the result of an imaginary gauge transformation on a Hermitian system \cite{hatano1996localization, hatano1997vortex, yao2018edge, lee2019anatomy, okuma2020topological}.  Take a 1D chain described by a Hamiltonian whose nonzero elements are $\mathcal{H}_{j+1,j}' = \mathcal{H}_{j,j+1}' = t_0$ (i.e., uniform, reciprocal, and Hermitian couplings) and $\mathcal{H}_{jj}' = V_j$.  Given an eigenstate $|\psi'\rangle = [\dots,\psi_j',\dots]^T$, we can perform a gauge transformation $\psi_j = e^{iBj} \psi_j'$, whereupon $|\psi\rangle$ is an eigenstate of the same energy for a Hamiltonian $\mathcal{H}^B$ whose nonzero elements are
\begin{align}
  \mathcal{H}_{j\pm1,j}^B &= t_0 e^{\pm iB} \\
  \mathcal{H}_{jj}^B &= V_j.
\end{align}
If we choose the imaginary gauge field $B = -i\kappa$, then $\mathcal{H}^B$ reduces to the Hatano-Nelson Hamiltonian \eqref{HN-1} with $t = t_0 \cosh\kappa$ and $\delta = t_0\sinh\kappa$.  For an infinite lattice, we should be wary of combining this gauge transformation with the usual Bloch decomposition, since it fails to preserve the form of the Bloch functions.  But if we consider a finite sample of the lattice with open  boundary conditions (i.e., Dirichlet-like boundary conditions with couplings set to zero at the lattice boundary), the argument implies that the eigenstates of $\mathcal{H}$ are the original eigenstates of $\mathcal{H}'$ modulated by exponential envelopes.  In particular, spatially-extended standing wave modes are transformed into exponentially localized modes called ``skin modes''.  Note that the gauge transformation preserves the eigenenergies, which therefore remain real.

The Hatano-Nelson model was originally developed to study the localization-delocalization transition in non-Hermitian systems, based on a disordered potential $\{V_j\}$ \cite{hatano1996localization}. For the remainder of this section, however, let us consider $V_j = 0$.  In this case, the model becomes translationally periodic, and we can study its Bloch eigenstates (which, as noted above, are \textit{not} the Bloch eigenstates of its gauge transformed counterpart).  By taking $\psi_j = \Psi \, e^{ikj}$, we can derive from Eq.~\eqref{HN-1} the complex energy band
\begin{equation}
  E_k = 2t\cos k + 2i\delta \sin k.
\end{equation}
This describes an elliptical trajectory in the complex plane, winding once as $k$ advances from 0 to $2\pi$, as shown in Fig.~\ref{fig:HNmodel}(a)--(b) by the colored curves labelled PBC (for ``periodic (i.e., Bloch-type) boundary conditions'').  On the other hand, the black dots in Fig.~\ref{fig:HNmodel}(a)--(b) show the eigenenergies for a finite sample of the lattice with open boundary conditions (labelled OBC).  These are real, as explained in the previous paragraph.  For the $\delta = 0$ case, shown in Fig.~\ref{fig:HNmodel}(a), the OBC eigenenergies are all zero since the Hamiltonian reduces to Eq.~\eqref{HO-3}, yielding an order-$N$ EP, where $N$ is the number of sites.  For the $\delta \ne 0$ case, the OBC eigenenergies occupy a finite span of the real line enclosed by (and nearly touching) the loop of PBC eigenenergies.

In Fig.~\ref{fig:HNmodel}(c)--(d), we plot the spatial distributions of the OBC eigenstates.  We see that they are indeed localized to the left boundary of the lattice (in subplot (c), all the wavefunctions are identical because of the order-$N$ EP).  The winding direction of the PBC eigenenergies, and the direction of localization, both depend on the model parameters.  For instance, if $t > 0$ and $\delta < 0$, the PBC eigenenergies would wind clockwise and the OBC eigenstates would be localized to the right boundary.

Another way to interpret the properties of the Hatano-Nelson model, which clarifies the relationship between the NHSE and band topology, was found by Okuma \textit{et al.}  Define an auxiliary Hamiltonian \cite{okuma2020topological}
\begin{equation}
\tilde{\mathcal{H}}(E)=\begin{pmatrix} 0 &\mathcal{H}-E\\ \mathcal{H}^\dagger-E^* & 0\end{pmatrix},
\label{HN-2}
\end{equation}
where $\mathcal{H}$ is the Hatano-Nelson Hamiltonian of Eq.~\eqref{HN-1}, and $E$ is an arbitrary complex parameter.  The auxiliary Hamiltonian $\tilde{\mathcal{H}}(E)$ is Hermitian and describes a ``ladder'' formed by two coupled 1D chains, as depicted in Fig.~\ref{fig:HNmodel}(e).  The nearest-neighbor couplings in the Hatano-Nelson lattice are translated into reciprocal and Hermitian inter-chain couplings (black solid and dashed lines).  The $E$ and $E^*$ terms translate into inter-chain couplings that are nonreciprocal but Hermitian (gray arrows).  If there is an eigenstate of $\tilde{\mathcal{H}}(E)$ with zero energy, i.e.,
\begin{equation}
  \tilde{\mathcal{H}}(E) \begin{pmatrix} \psi_a \\ \psi_b \end{pmatrix}=0,
  \label{HN-3}
\end{equation}
then $(\mathcal{H}-E) \psi_b=0$, meaning that $\psi_b$ is an eigenstate of the Hatano-Nelson lattice with eigenvalue $E$.

Importantly, $\tilde{\mathcal{H}}(E)$ satisfies the chiral symmetry $\{\hat{C}, \tilde{\mathcal{H}}(E)\} = 0$ for every $E$ (see Sec.~\ref{sec:Symmetry}), where $\hat{C}=\sigma_3 \otimes I_N$, $I_N$ is the $N\times N$ identity matrix, and $N$ is the number of sites in the Hatano-Nelson lattice.   This implies that, like the SSH lattice described in Sec.~\ref{sec:wavetopology}, the bandstructure can be topologically characterized by a Zak phase quantized to $0$ or $\pi$ \cite{hasan2010colloquium, BansilReview2016, Chiu2016}, with a nonzero value implying the existence of topological zero modes for the finite (OBC) lattice.  Taking the $\mathcal{H}$ in Eq.~\eqref{HN-2} to be the momentum space Hamiltonian of the Hatano-Nelson model, its Zak phase can be shown to be $\pi \cdot W(E)$, where $W(E)$ is the number of times the PBC eigenenergies wind around the reference energy $E$ as the quasimomentum $k$ advances across the Brillouin zone.  Thus, for each OBC eigenenergy $E$ shown in Fig.~\ref{fig:HNmodel}(a)--(b), for which $W(E) = 1$, $\tilde{\mathcal{H}}(E)$ has a nonzero Zak phase, so its finite-lattice counterpart exhibits zero modes, so the corresponding eigenstate of $\tilde{\mathcal{H}}(E)$ is a topological boundary state.  This explains why every eigenstate of $\mathcal{H}$ is localized.

The NHSE can also be observed in 1D lattices other than the Hatano-Nelson model.  A concept called the ``generalized Brillouin zone'' provides a general way to explain how the NHSE arises in 1D models \cite{yao2018edge, yokomizo2019non, yang2020non, zhang2020correspondence, guo2021exact, wu2022connections}.  This involves replacing $e^{ik}$ with $z$ in the bulk Hamiltonian, and expressing the bulk eigenfunctions as polynomials.  The $|z| = 1$ case corresponds to the standard Brillouin zone, whereas the $|z|\ne 1$ behavior can be used to infer whether the model exhibits the NHSE.  Notably, this technique applies to 1D lattices beyond the Hatano-Nelson model, including those with different (or no) symmetries.  A pedagogical treatment is given in Ref.~\onlinecite{zhang2020correspondence}.

The NHSE mostly occurs in, and is associated with, lattices with nonreciprocal couplings.  However, this is not always the case, and the NHSE can also be found in certain lattices with entirely recipocal couplings, usually as a result of additional lattice symmetries.  For example, Lee \textit{et al.}~introduced a 1D lattice with only on-site gain/loss that exhibits the NHSE \cite{lee2016anomalous}. Another interesting phenomenon is called the Z$_2$ NHSE, whereby two NHSEs are simultaneously in effect on opposite boundaries \cite{okuma2020topological}.

\begin{figure*}
\centering
\includegraphics[width=0.96\textwidth]{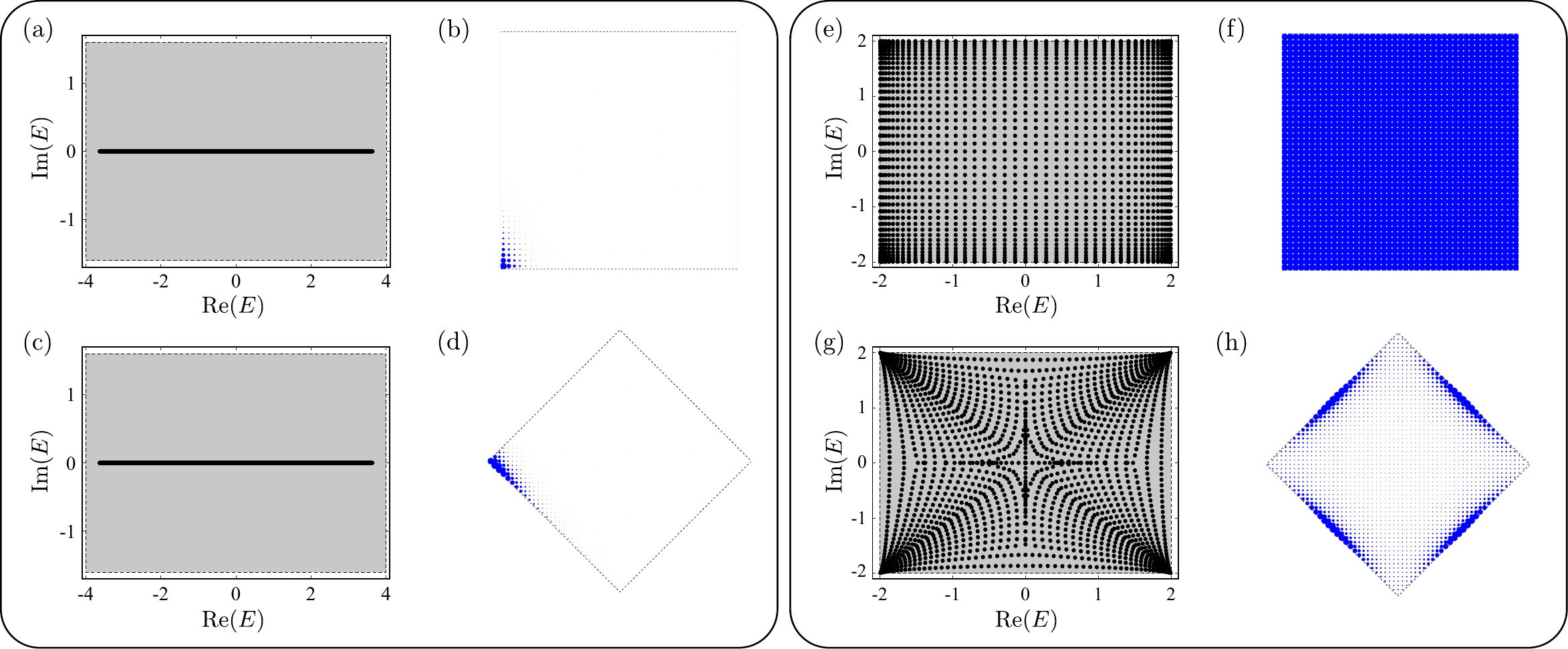}
\caption{Non-Hermitian skin effect in 2D lattices.  (a) Eigenenergies for the infinite lattice (gray area) and for a finite lattice truncated to a square with $N = 1849$ sites, for parameters $t_x=1$, $t_y=1$, $\delta_x=0.5$ and $\delta_y=0.3$.  (b) Spatial distribution of the total intensity $\mathcal{I}(x,y)=\sum_{n=1}^N |\langle x,y|n\rangle|^2$, where $|n\rangle$ represent the normalized $n$-th eigenstate and $(x,y)$ are lattice coordinates.  The radius of each blue circle is proportional to $\mathcal{I}(x,y)$.  (c)--(d)  Similar to (a)--(b), but with the finite lattice truncated to a diamond (i.e., a square oriented 45 degrees to the lattice axes) with $N = 1861$ sites.  (e)--(h) Similar to (a)--(d), with the parameters $t_x=1$, $t_y=0$, $\delta_x=0$ and $\delta_y=1$. }
\label{fig:NHSE2D}
\end{figure*}

\subsection{Non-Hermitian Skin Effect in Higher Dimensions}
\label{sec:nhse2d}

The NHSE also occurs in 2D or higher-dimensional lattices.  For example, by stacking the 1D Hatano-Nelson model along the $y$ axis, we can construct a 2D lattice whose Hamiltonian has nonzero elements
\begin{align}
  \begin{aligned}
    \mathcal{H}_{i,j;i-1,j} &= t_x-\delta_x \\
    \mathcal{H}_{i,j;i+1,j} &= t_x+\delta_x \\
    \mathcal{H}_{i,j;i,j-1} &= t_y-\delta_y \\
    \mathcal{H}_{i,j;i,j+1} &= t_y+\delta_y,
  \end{aligned}
  \label{HN-8}
\end{align}
where lattice index $(i,j)$ refers to the $i$-th column ($x$ axis) and $j$-th row ($y$ axis).  The 2D lattice has couplings $t_x \pm \delta_x$ and $t_y \pm \delta_y$ along $x$ and $y$ respectively, which are nonreciprocal when $\delta_x$ or $\delta_y$ are nonzero. 

Consider the exemplary case $t_x = t_y=1$, $\delta_x=0.5$ and $\delta_y=0.3$.  In Fig.~\ref{fig:NHSE2D}(a), we plot the spectrum of the infinite lattice (PBC), which occupy the area shaded in gray, and the spectrum of a finite square-shaped sample (OBC), which occupies a finite segment of the real line, as shown by the black dots.  In Fig.~\ref{fig:NHSE2D}(b), we plot the spatial distribution of the total intensity $\mathcal{I}(x,y)=\sum_n^N |\langle x,y|n\rangle|^2$, where $|n\rangle$ is the $n$-th eigenstate, $N$ is the total number of sites, and $(x,y)$ are 2D lattice coordinates.  Evidently, the eigenstates are localized to the bottom-left corner of the sample.  If we now change the finite lattice to a diamond shape, the spectrum and intensity distribution are as shown in Fig.~\ref{fig:NHSE2D}(c)--(d); the OBC eigenstates still have real eigenenergies, and are now localized on the left corner of the diamond.  Other finite lattice shapes exhibit the NHSE in a similar fashion.  We can understand this behavior as a consequence of stacking the 1D Hatano-Nelson model, causing the nonreciprocal couplings to ``push'' the eigenstates to one side of the lattice along each spatial dimension.

Next, consider the case $t_x=1$, $t_y=0$, $\delta_x=0$ and $\delta_y=1$.  As before, the PBC eigenenergies occupy a spectral area, plotted in gray in Fig.~\ref{fig:NHSE2D}(e) and (g).  On the other hand, the OBC eigenenergies for both the square and diamond lattices are no longer constrained to the real line, and occupy approximately the same area as the PBC eigenenergies, as shown by the black dots.  However, the total intensity plots, in Fig.~\ref{fig:NHSE2D}(f) and (h), reveal a striking difference between the two samples: for the square-shaped lattice, no NHSE present, but in the diamond-shaped lattice the eigenstates are localized to the four edges of the diamond.  Because the manifestation of the NHSE depends in this case on the shape of the lattice, it is referred to as a ``geometry-dependent'' NHSE; the preceding case, in which the NHSE is shape-independent, is called the ``strong'' NHSE \cite{zhang2022universal}.

At present, whether the NHSE in 2D and higher dimensions can be rigorously linked to band or spectral topology remains an unsettled question.  It is possible, however, to offer some intuitive connections between the presence or absence of the NHSE and the complex spectrum \cite{zhang2022universal}.  First consider a 1D Hermitian lattice: in a finite lattice, each eigenstate of energy $E$ should be a standing wave formed by a superposition of two or more Bloch waves, with different real $k$ and the same $E$ (the coefficients of the superposition are determined by the OBC on the total wavefunction).  As the Bloch band energies are periodic modulo the Brillouin zone, there are always at least two $k$ for each $E$ within a band (away from band extrema).  However, this is not the case for a non-Hermitian 1D lattice.  For example, we saw in Fig.~\ref{fig:HNmodel}(a) that the Hatano-Nelson model has a spectral loop of winding number 1.  Each complex $E$ on the loop maps to only one $k$, so a standing wave cannot be formed.

Consider how this argument extends to 2D or higher dimensions.  The formation of a standing wave in a finite lattice of arbitrary shape may now require an infinite set of Bloch states, all having the same energy $E$ (which may be complex) and real $k(C)$, where $C$ is some trajectory in momentum space.  In the Hermitian case, each band is a real spectral arc (see Sec.~\ref{sec:Gap}), so the mapping from 2D momentum space to a 1D arc guarantees that each $E$ within the band maps to infinitely many $k$.  This also applies to spectral arc bands in non-Hermitian lattices.  But for a non-Hermitian band that forms a spectral area, each $E$ in the 2D spectral area might map to only a finite set of $k$ in the 2D momentum space.  For some shapes, it might still be possible to satisfy the OBC using only finitely many real $k$, whereas for other shapes the formation of standing waves is obstructed.  This leads to the geometry-dependent NHSE \cite{zhang2022universal}.
 
Recent theoretical work has uncovered a couple of variants of the NHSE that occur in 2D or higher dimensions.  One is called the higher-order NHSE \cite{kawabata2020higher}.  Normally, a $D$-dimensional lattice exhibiting the NHSE has $\mathcal{O}(L^D)$ skin modes, where $L$ is the system size.  In the higher-order NHSE, the scaling changes to $\mathcal{O}(L^{d})$ where $d<D$ \cite{lee2019hybrid, kawabata2020higher, okugawa2020second, kim2021disorder, zou2021observation, zhang2021observation, fu2021non}. Another type of NHSE, called the reciprocal NHSE, has been found in 2D Hamiltonians satisfying the reciprocity symmetry $\mathcal{H}(k_x,k_y)=H^T(-k_x,-k_y)$.  When OBC is imposed along $x$ and PBC along $y$, the lattice exhibits skin modes which are localized at opposite boundaries for $\pm k_y$ \cite{hofmann2020reciprocal}.



\section{Experimental platforms}
\label{sec:platform}

In this section, we will survey the various photonics platforms for implementing non-Hermitian photonic lattices, and discuss their different advantages and disadvantages.

\begin{figure*}
\centering
\includegraphics[width=0.96\textwidth]{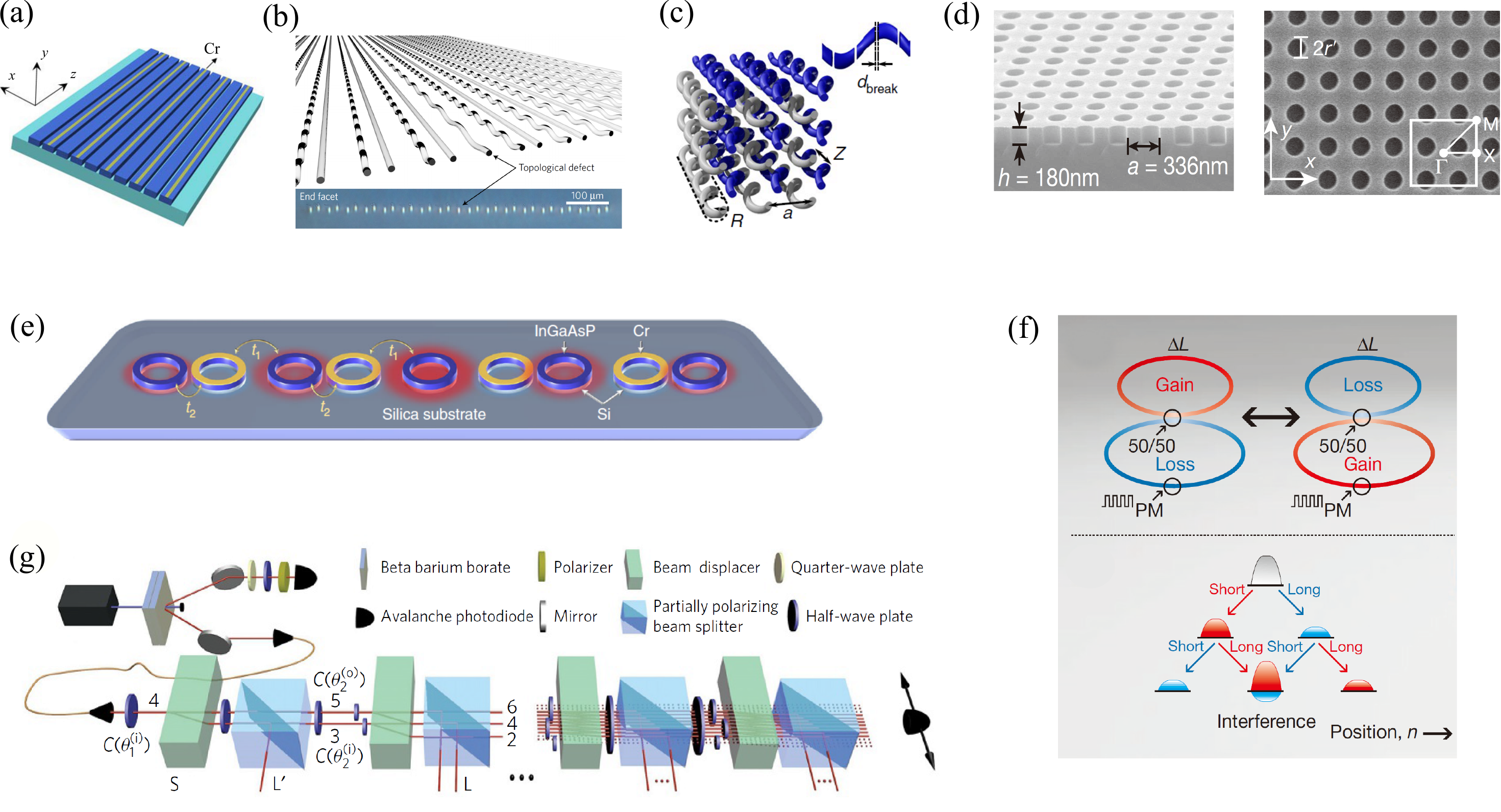}
\caption{Experimental platform for implementing non-Hermitian photonic lattices. (a) Schematic of a 1D PT symmetric waveguide array that was fabricated on a silicon-on-insulator chip with chrome strips to implement losses on one sublattice.  Reproduced from Ref. \cite{xu2016experimental}. (b) Schematic of a 1D PT symmetric waveguide array based on laser-written optical waveguides in fused silica, with radiative losses induced by ``wiggling'' one set of waveguides. Reproduced with permission from Springer Nature from Ref. \cite{weimann2017topologically}. (c) Schematic of a 2D non-Hermitian lattice based on laser-written waveguides in borosilicate glass.  The waveguides on one sublattice are fabricated with small breaks (see inset) to enhance their radiative losses, in order to produce a bandstructure with an exceptional ring.  Reproduced with permission from Springer Nature from Ref. \cite{cerjan2019}. (d)  A 2D photonic crystal formed by a square array of holes etched into a silicon nitride slab, designed to support exceptional points in the bandstructure. Reproduced with permission from Springer Nature from Ref.~\cite{zhen2015spawning}.  (e) Schematic of a 1D cSSH lattice formed by microring resonators, with gain provided by a optically pumped InGaAsP on the rings in one sublattice.  Reproduced from Ref.~\cite{zhao2018topological}.  (f) Schematic of coupled fiber loops with differential gain and loss.  Pulses propagating in these loops simulate wavefunction evolution in a time-modulated 1D PT symmetric lattice. Reproduced with permission from Springer Nature from Ref.~\cite{regensburger2012parity}.  (g) Experimental setup for a quantum walk experiment that simulates wavefunction evolution in a PT symmetric lattice.  Reproduced with permission from Springer Nature from Ref.~\cite{xiao2017observation}.
  }
\label{fig:Experiments}
\end{figure*}

\subsection{Optical Waveguide Arrays}
\label{sec:waveguide_arrays}

Optical waveguide arrays have long been leading platforms for studying non-Hermitian lattices \cite{guo2009observation, doppler2016dynamically, hassan2017dynamically, yoon2018time, zhang2019dynamically, shu2022fast, schumer2022topological}. When light propagates down an array of parallel waveguides, its envelope obeys an evolution equation formally equivalent to a time-dependent Schr\"odinger equation \cite{lax1975maxwell, makris2008beam, rechtsman2013}, where the propagation distance along the waveguide axis, $z$, maps to the dimension of time $t$:
\begin{equation}
i\frac{\partial{\psi(\mathbf{r}_\perp,z)}}{\partial{z}} = \left[- \frac{1}{2k_0}\nabla^2_\perp -\frac{k_0}{n_0}\delta_n(\mathbf{r}_\perp,z)\right]\psi(\mathbf{r}_\perp,z).
\label{progation}
\end{equation}
Here, $\mathbf{r}_\perp = (x,y)$ denotes the coordinates transverse to the waveguide axis, $n_0$ is the background refractive index, $\delta_n(\mathbf{r}_\perp,z)$ is the additional contribution to the refractive index associated with the waveguides (which can be complex), and $\psi(\mathbf{r}_\perp,z)$ is the electric field envelope.  The physical electric field is given by $E(\mathbf{r}_\perp,z,t) = \psi(\mathbf{r}_\perp,z) \exp\left[i(k_0z-\omega_0t)\right]$, where $\omega_0$ is the operating frequency and $k_0$ is the wavenumber in the background medium.  Eq.~\eqref{progation} can be derived from Maxwell's equations within the ``paraxial limit'', under the assumption that  the waveguides' cross sectional profiles vary over length scales $\Delta z$ much longer than the optical wavelength $2\pi/k_0$ \cite{lax1975maxwell}.  Thus, an array of waveguides with $\delta_n > 0$ maps to an array of potential wells defined in a 2D space.  Under the tight-binding approximation, this can be used to define a 1D or 2D lattice \cite{makris2008beam, rechtsman2013}.  As light diffracts through the waveguide array, the guided optical modes essentially hop between the waveguides (or lattice sites), without leaking into unguided free-space modes; the lack of such leakage manifests in the Hermiticity of Eq.~\eqref{progation} when $n_0, \delta_n \in \mathbb{R}$.

There are two common ways to implement waveguide arrays.  The first is to fabricate ridge waveguides onto a planar substrate, as shown in Fig.~\ref{fig:Experiments}(a).  This approach can make use of well-developed nanofabrication technologies, and probing techniques such as the use of scanning near-field optical microscopy to measure the field distribution; however, it is generally limited to realizing 1D lattices, with one transverse spatial dimension and the axial direction mapping to time \cite{lederer2008discrete}.  The second method is to laser-write waveguides directly into a transparent medium, as shown in Fig.~\ref{fig:Experiments}(b); this allows for the implementation of not only 1D but also 2D lattices, with highly flexible geometries \cite{meany2015laser}.  Both types of waveguide arrays have been extensively used to study Hermitian phenomena such as Anderson localization \cite{martin2011anderson} and photonic topological edge states \cite{rechtsman2013}.

Non-Hermiticity is typically introduced into waveguide arrays via deliberately placed losses (``loss engineering''), corresponding to imaginary on-site lattice potentials.  For instance, lossy materials can be incorporated into the waveguides \cite{guo2009observation, xu2016experimental, pan2018photonic, song2019breakup}, as was done in a non-Hermitian waveguide array fabricated on a silicon-on insulator chip by Xu \textit{et al.}~\cite{xu2016experimental}, depicted in Fig.~\ref{fig:Experiments}(a).  Loss-inducing chrome strips were deposited on half of the waveguides, so that, up to an average overall decay rate, the structure served as a realization of a PT symmetric 1D lattice (see Sec.~\ref{sec:pt}).

Loss engineering can also be accomplished by manipulating the waveguides' radiative losses \cite{weimann2017topologically, cerjan2019}.  For example, Weimann \textit{et al.}~realized a non-Hermitian 1D SSH lattice in a laser-written waveguide array, with radiative losses enhanced in one set of waveguides by giving them a sinusoidal modulation, as shown in Fig.~\ref{fig:Experiments}(b).  Cerjan \textit{et al.}~introduced radiative losses by fabricating small breaks along the waveguides, as shown in Fig.~\ref{fig:Experiments}(c) \cite{cerjan2019}, and hence implemented a 2D lattice with an exceptional ring in its bandstructure (see Sec.~\ref{sec:Lattices2D}).  Recently, Xia \textit{et al.} have extended such non-Hermitian lattices into the nonlinear regime by writing waveguides into photorefractive crystals, and demonstrating the use of optical nonlinearity to drive a non-Hermitian phase transition \cite{xia2021nonlinear}.

When investigating theoretical models that posit both gain and loss, it is common to use loss engineering to bypass implementing gain (which is often inconvenient in experiments).  This is achieved by imposing an overall loss level on the system.  For example, in a PT symmetric Hamiltonian [such as Eq.~\eqref{H2by2} with $\delta = 0$], one can replace the balanced gain/loss terms $\pm i\gamma$ with two unequal loss terms, $i\gamma_1$ and $i\gamma_2$.  In the resulting non-Hermitian dynamics, the state vector $|\psi'(z)\rangle$ is
\begin{equation}
  |\psi'(z)\rangle = \exp\left(-\frac{(\gamma_1+\gamma_2)t}{2}\right) |\psi(z)\rangle,
\end{equation}
where $|\psi(z)\rangle$ is the state vector for the original PT symmetric system with $\gamma=|\gamma_1-\gamma_2|/2$ \cite{guo2009observation, lawrence2014manifestation, doppler2016dynamically}.

However, actual gain can also be incorporated into waveguide arrays \cite{ruter2010observation, schumer2022topological}.  R\"uter \textit{et al.}, in one of the earliest works on PT symmetric photonics, demonstrated the optical pumping of Fe-doped LiNbO$_3$ waveguides to achieve positive gain \cite{ruter2010observation}.  Recently, Li \textit{et al.}~fabricated a cSSH waveguide array (Sec.~\ref{sec:Lattices1D}) with gain provided by electrically pumped quantum dots, allowing for a tunable laser device based on cSSH boundary states \cite{li2022electrically}.

As experimental platforms, waveguide arrays have two significant advantages.  First, because the axial distance $z$ maps to time, the time-evolution of the wavefuntion can be relatively easily measured (e.g., by imaging the intensity distribution at the end-facet of a laser-written waveguide array of fixed length $\delta z$ \cite{rechtsman2013, meany2015laser, weimann2017topologically}).  This has been used, for example, in studying the dynamical effects of non-Hermitian topological defect states \cite{weimann2017topologically}.  Second, certain ``time''-dependent Hamiltonians can be implemented relatively easily, by introducing slow modulations to the cross sectional features of the waveguide array.  For instance, curved waveguides have been used to study non-Hermitian Bloch oscillations \cite{xu2016experimental}.  Some limitations of waveguide arrays include the difficulty of fabricating very large-scale lattices, the relative difficulty of probing the bandstructure (because exciting the lattice at a given ``frequency'' translates into specifying a transverse wavenumber $k_z$), and the impossibility (absent additional tricks) of implementing 3D lattices.


The paraxial wave equation \eqref{progation} explicitly assumes that light travels in a single direction along the waveguides, without backscattering.  If both forward and backward propagation is allowed, the variation along $z$ is described not by an evolution matrix, but by a transfer matrix generated by a non-Hermitian matrix.  This might be an interesting way to access non-Hermitian Hamiltonians without requiring actual gain or loss \cite{chen2017pseudo, nada2017theory}, though one that has been little explored.

\subsection{Photonic Crystals}

Photonic crystals are a promising platform for studying the physics of non-Hermitian lattices, with non-Hermiticity typically introduced via loss engineering \cite{zhen2015spawning, wu2020vector, zhou2018observation}.  Compared to waveguide arrays, photonic crystals are highly compact, and can be fabricated in 3D geometries; moreover, photonic crystal bandstructures can be probed experimentally using far-field measurements, whereas their internal wavefunctions are very difficult to map out.  Another notable difference is that photonic crystals cannot be directly described by tight-binding models like those that have guided our discussion thus far; however, tight-binding models can still serve as qualitative guides.

Zhen \textit{et al.}~fabricated a two-dimensional photonic crystal slab, depicted in Fig.~\ref{fig:Experiments}(d), whose bandstructure contains an exceptional ring \cite{zhen2015spawning}.  The photonic crystal consisted of a square pattern of air holes etched into in a Si$_3$N$_4$ slab, with losses stemming from mode-dependent out-of-plane radiative loss (e.g., dipole modes out-couple while quadrupole modes do not).  By tuning the lattice parameters, they set up two dipole modes and one quadrupole mode to have eigenfrequencies with the same real part but different imaginary parts, enabling the formation of an exceptional ring in the bandstructure.  Based on a similar idea, Zhou \textit{et al.}~designed a rhombic lattice with elliptical air holes and observed pairs of EPs connected by ``bulk fermi arc'' modes \cite{zhou2018observation}.

Recently, Zhong \textit{et al.}~showed theoretically that a photonic crystal incorporating lossy materials can exhibit spectral loops and the NHSE \cite{zhong2021nontrivial}.  Notably, the design utilizes purely local losses, without any evident connection to the nonreciprocal couplings commonly associated with the NHSE in tight-binding models (see Sec.~\ref{sec:nhse}).  This design has not yet been realized experimentally.

\subsection{Micro-Resonator Arrays}

Arrays of micro-resonators can be used to realize 1D and 2D photonic lattices.  Such designs have been extensively used in the study of non-Hermitian photonics over the past decade \cite{peng2014parity, chang2014parity, feng2014single, hodaei2014parity, brandstetter2014reversing, hodaei2016single, peng2016chiral, miao2016orbital, peng2016chiral, liu2017integrated, parto2018edge, lafalce2019robust, wang2020electromagnetically, zhang2020tunable, zhang2020synthetic}.  They commonly employ ring resonators, which can achieve very high Q factors.
For example, Zhao \textit{et al.}~realized a cSSH lattice (Sec.~\ref{sec:Lattices1D}) using a microring array, as depicted in Fig.~\ref{fig:Experiments}(e) \cite{zhao2018topological}.  Each microring contained either InGaAsP, to induce gain under optical pumping, or a chrome layer to induce losses.  Robust single-mode lasing based on a cSSH boundary state was observed.  These authors also fabricated a 2D lattice based on similar design principles, using it to demonstrate the dynamical control over topological boundary states via optical pumping \cite{zhao2019non}.

Microring arrays can also be used to implement nonreciprocal couplings, such as those that can give rise to the NHSE (see Sec.~\ref{sec:nhse}).  This is accomplished by placing an off-resonant ``link ring'' between each neighboring pair of primary resonators \cite{Leykam2018}.  Gain and loss are incorporated into opposite arms of the link ring, and the differential gain/loss when passing through in opposite directions acts as a nonreciprocal coupling \cite{zhu2020photonic, lin2021steering}.  This technique has recently been demonstrated experimentally in acoustics \cite{zhang2021observation, gao2022anomalous}, followed by a photonics implementation \cite{liu2022complex}.

Aside from microring resonators, arrays can also be constructed from other kinds of high-Q resonators \cite{dembowski2001experimental, poli2015selective, rao2021controlling, reisner2021microwave}.  For example, Poli~\textit{et al.}~implemented the cSSH lattice in the microwave regime using cylindrical dielectric resonators, with additional losses achieved by depositing microwave-absorbing material on some of the cylinders \cite{poli2015selective}.  A significant advantage of microwave experiments, compared with those done at optical frequencies, is that the amplitude and phase of the wavefunction can be easily measured experimentally. 

\subsection{Optical Fiber Loops}

Optical fibers provide an interesting way to study the physics of non-Hermitian lattices \cite{regensburger2012parity, regensburger2013observation, wimmer2015observationBloch, wimmer2015observation, jahromi2017statistical, weidemann2020topological, wang2021generating, bergman2021observation, steinfurth2022observation, weidemann2022topological, nasari2022observation}.  Using established technologies for controlling pulses in optical fibers (e.g., fiber-coupled electro-optic modulators), it is possible to simulate lattices features that are hard to achieve on other platforms, such as time-dependent gain/loss modulations.

Regensburger \textit{et al.}~realized a synthetic PT symmetric lattice using the setup shown in Fig.~\ref{fig:Experiments}(f) \cite{regensburger2012parity}.  Conceptually, this and related experiments rely on a formal equivalence between the circulation of optical pulses in a set of coupled loops and the time evolution of a wavefunction in a discrete lattice.  Here, the two sublattices of a PT symmetric 1D lattice were represented by two fiber loops connected by a 50/50 coupler, and non-Hermiticity was introduced by using acousto-optic modulators to add gain to the pulses within one loop.  The same group also demonstrated the existence of PT symmetric solitons by utilizing the optical nonlinearity in the fibers \cite{wimmer2015observation}.  Recently, the NHSE was observed in a fiber loop setup by using direction-dependent amplification and attenuation to simulate nonreciprocal couplings \cite{weidemann2020topological}.

A different approach was taken by Wang \textit{et al.}, who used a single fiber loop with amplitude and phase modulations to implement a synthetic lattice with tunable nonreciprocal and/or non-nearest-neighbor couplings \cite{wang2021generating}.  These authors were able to demonstrate active control over the spectral winding of the synthetic bands.

\subsection{Other Platforms}

Non-Hermitian photonic lattices can be simulated using ``quantum walk'' experiments, which involve single photons passing through sequences of optical elements \cite{xiao2017observation, zhan2017detecting, xiao2019observation, xiao2020non, xiao2021observation, wang2021simulating, lin2022observation}.  For example, Xiao \textit{et al.}~designed a PT symmetric quantum walk using the setup shown in Fig.~\ref{fig:Experiments}(g) \cite{xiao2017observation}.  Different losses in the two sublattices were implemented simply by the placement of a partially polarizing beam splitter within the apparatus.  The NHSE was later observed by the same group using a similar setup \cite{xiao2020non}.

Non-Hermitian photonic lattices can also be implemented in exciton-polariton systems \cite{gao2015observation, baboux2016bosonic, gao2018chiral}, optomechanical systems \cite{xu2016topological}, and ultracold atoms lattices \cite{zhang2016observation, li2019observation}, which all lie outside the scope of our present discussion.

\section{Outlook}
\label{sec:outlook}

In this tutorial, we have surveyed the properties of non-Hermitian systems with discrete translational symmetry, and discussed how they can be realized on photonics-based experimental platforms.  This has been and continues to be a rapidly-evolving field of theoretical and experimental research; indeed, we have been forced to omit some interesting sub-topics, in the interest of space.  In the future, there are several areas in which important advancements may yet be made.

The first concerns the practical question of how to reliably perform gain/loss engineering on photonic lattices, in order to realize Hamiltonians with more complicated forms of non-Hermiticity.  To date, demonstrations of gain/loss engineering on various photonic platforms have tended to focus on PT symmetry, which can be regarded as one of the simplest nontrivial forms of non-Hermiticity.  As we have seen, there are other interesting and consequential non-Hermitian symmetries (Sec.~\ref{sec:Symmetry}), but these are often more challenging to implement, e.g.~because they involve more complicated distributions of gain/loss.  Realizing and controlling nonreciprocal couplings, which are important for phenomena such as the non-Hermitian skin effect (Sec.~\ref{sec:nhse}), is also an open practical problem on several platforms, such as microresonator arrays.  Such technical advances would also open the door to applications such as light funneling \cite{feng2017non, zhao2018parity, miri2019exceptional, weidemann2020topological, parto2021non}.

Non-Hermitian band topology is a rapidly developing topic of research (Sec.~\ref{sec:topology}), with many important open questions.  We anticipate further progress on formulating the topological characterization of non-Hermitian bandstructures, understanding the precise nature of topological bulk-boundary correspondence in non-Hermitian lattices, and experimentally implementing new non-Hermitian topological phenomena.  It will be particularly interesting to see if the recent categorization of non-Hermitian topological insulators \cite{kawabata2019symmetry} will prove useful for guiding the development of new photonic lattices, which could exploit unique features of non-Hermitian topological phases not found in the Hermitian regime.  Also, the relationship between the non-Hermitian skin effect and topology, though apparently well-established in 1D, is still in need of a better level of theoretical understanding in 2D and higher dimensions (Sec.~\ref{sec:nhse2d}).

While we have focused on crystalline lattices, there are many interesting phenenomena associated with disordered lattices, or lattices where the crystalline symmetry is violated by defects \cite{mermin1979topological}.  For instance, topological lattice defects can alter the non-Hermitian skin effect \cite{sun2021geometric, schindler2021dislocation, bhargava2021non}, or provide a way to probe subtle aspects of non-Hermitian band topology \cite{schindler2021dislocation}.


Another area that is worth exploring is the implications of non-Hermiticity for quantum photonic lattices.  In this tutorial, we have focused on the ``single-particle'' picture, which corresponds to the regime of classical photonics; we have not covered quantum effects such as multi-photon dynamics, entanglement, and squeezing.  Non-Hermitian phenomena can be introduced into the quantum regime via several interesting avenues \cite{song2019non, yi2020non, longhi2020unraveling, flynn2021topology, wang2021giant, mcdonald2022nonequilibrium, faugno2022interaction}.  For example, certain particle non-conserving Hamiltonians, describing parametrically driven nonlinear systems \cite{boyd2020nonlinear}, can be mapped to non-Hermitian Hamiltonians via the Bogoliubov-de Gennes transformation \cite{rossignoli2005complex, clerk2010introduction, caves2012quantum, galilo2015selective, engelhardt2016topological}, which can be used to access phenomena such as non-Hermitian topological boundary states and the non-Hermitian skin effect \cite{barnett2013edge, peano2016topological, mcdonald2018phase, wang2022amplification}.  The experimental implementation of quantum photonic lattices exhibiting non-Hermitian phenomena, though highly challenging, would provide new opportunities for realizing non-Hermitian bandstructures and exploiting their special properties.\\

\noindent \textbf{Funding.} This work was supported by the Singapore MOE Academic Research Fund Tier~3 Grant MOE2016-T3-1-006, and by the National Research Foundation Competitive Research Programs NRF-CRP23-2019-0005 and NRF-CRP23-2019-0007. 

\noindent \textbf{Disclosures.} The authors declare no conflicts of interest. 

\noindent \textbf{Data availability.} The data for all plots, except those excerpted from other works, is available upon request.

\bibliography{citepaper}
\end{document}